\documentclass[11pt,a4paper]{article}
\pdfoutput=1

\usepackage{jheppub}
\usepackage{amsmath}
\usepackage{amssymb}
\usepackage{graphics}
\usepackage[active]{srcltx}
\usepackage{pdfsync}
\usepackage{shuffle}
\usepackage{slashed}
\usepackage{hyperref}
\usepackage{subfigure}

\newcommand{\bea}{\begin{eqnarray}}
\newcommand{\eea}{\end{eqnarray}}
\newcommand{\nn}{\nonumber \\}

\def\W #1{\widetilde{#1}}

\def\eref#1{(\ref{#1})}

\def\a{{\alpha}}

\def\b{{\beta}}

\allowdisplaybreaks

\title{Hidden zeros for higher-derivative YM and GR amplitudes at tree-level}
\author[a]{Kang Zhou}

\affiliation[a]{Center for Gravitation and Cosmology, College of Physical Science and Technology, Yangzhou University,\\
No.180, Siwangting Road, Yangzhou, 225009, P.R. China}

\emailAdd{zhoukang@yzu.edu.cn}

\date{\today}
\abstract{We extend the recently discovered phenomenon of hidden zeros to tree amplitudes for Yang-Mills (YM) and general relativity (GR) theories with higher-derivative interactions. This includes gluon amplitudes with a single insertion of the local $F^3$ operator, as well as graviton amplitudes at sub-leading and sub-sub-leading orders in the low-energy expansion of bosonic closed string amplitudes---referred to as $R^2$ and $R^3$ amplitudes, respectively.
The kinematic condition for hidden zeros leads to unavoidable propagator singularities in unordered graviton amplitudes. We investigate in detail the systematic cancellation of these divergences, which resolves ambiguities in the proof of hidden zeros. Our approach is based on universal expansions that express tree amplitudes as linear combinations of bi-adjoint scalar amplitudes.

}

\keywords{Scattering Amplitudes, Hidden Zero, Expansion}

\begin{document}

\maketitle \flushbottom

\section{Introduction}
\label{sec-intro}

Recent advances in the study of the S-matrix have uncovered a range of remarkable properties in scattering amplitudes that are not apparent within the traditional framework of Feynman rules. These discoveries have been twofold: on one hand, they revealed profound underlying properties of scattering amplitudes, such as color-kinematics duality and the double-copy construction \cite{Bern:2008qj,Bern:2010ue,Bern:2019prr}; on the other hand, they led to the development of new formulations that redefine amplitudes themselves, including the Cachazo-He-Yuan (CHY) formalism \cite{Cachazo:2013hca,Cachazo:2013iea,Cachazo:2014nsa,Cachazo:2014xea}, along with various geometric and combinatorial representations \cite{Arkani-Hamed:2017mur,Arkani-Hamed:2023lbd,Arkani-Hamed:2023mvg,Arkani-Hamed:2023jry,Arkani-Hamed:2024nhp,Arkani-Hamed:2024yvu,
Arkani-Hamed:2024tzl,Arkani-Hamed:2024nzc,Arkani-Hamed:2024pzc}.
A further significant discovery in this lineage of amplitude properties is the phenomenon of ``hidden zeros" in tree amplitudes and their accompanying factorization behaviors, as unveiled in \cite{Arkani-Hamed:2023swr}. Using kinematic mesh and stringy curve integral techniques, the authors demonstrated that tree amplitudes in theories such as ${\rm Tr}(\phi^3)$, the non-linear sigma model (NLSM), and Yang-Mills (YM) vanish on certain special loci in kinematic space. When a vanishing Mandelstam variable in the hidden zero kinematics is turned on, each amplitude factorizes into three amputated currents. Subsequent works in \cite{Rodina:2024yfc,Bartsch:2024amu,Li:2024qfp,Zhang:2024iun,Zhou:2024ddy,Zhang:2024efe,Feng:2025ofq,Cao:2025hio,Zhang:2025zjx,Huang:2025blb,
Backus:2025hpn} have further explored these hidden zeros. These studies have provided new perspectives on this phenomenon and extended the investigation to unordered tree amplitudes---in theories such as the special Galileon (SG), Dirac-Born-Infeld (DBI), general relativity (GR), and form factors, and the loop-level. In addition, the associated factorization properties, which occur without taking residues at any pole, have stimulated extensive research, as documented in \cite{Cao:2024gln,Arkani-Hamed:2024fyd,Cao:2024qpp,Zhang:2024iun,Zhou:2024ddy,Zhang:2024efe,Feng:2025ofq,Feng:2025dci,Cao:2025hio,Zhang:2025zjx}.

Hidden zeros are of considerable significance. From a theoretical perspective, it is both interesting and important to investigate whether the hidden zeros, in conjunction with the factorization on physical poles, is sufficient to uniquely determine scattering amplitudes. On the practical side, hidden zeros provide a novel constraint that facilitates the construction of amplitudes. For example, as demonstrated in \cite{Li:2025suo}, they can be employed to establish a new on-shell recursion relation for tree NLSM amplitudes, thereby circumventing the challenges posed by boundary terms.

Given the significance of hidden zeros, it is natural to investigate whether they also exist in tree amplitudes of other physical theories. In this work, we establish their existence in YM and GR amplitudes that include special higher-derivative interactions. The higher-derivative YM amplitudes studied in this paper are gluon amplitudes incorporating a single insertion of of the local $F^3$ operator. This operator corresponds to the leading correction to usual YM theory in the low-energy effective action of bosonic open string theory \cite{Polchinski:1998rq}. Furthermore, it is regarded as a potential signature of deviations in gluon interactions from conventional QCD predictions, possibly originating from new physics \cite{Simmons:1989zs,Simmons:1990dh,Cho:1993eu}. On the gravitational side, the higher-derivative GR amplitudes under investigation are graviton amplitudes that appear as sub-leading and sub-sub-leading terms in the low-energy expansion of bosonic closed string amplitudes. These are commonly referred to as $R^2$ and $R^3$ amplitudes, respectively. The above tree amplitudes, which typically arise in effective field theories, capture low-energy manifestations of unknown high-energy physics As will be shown in this paper, all such higher-derivative amplitudes exhibit their hidden zeros.

We expose the hidden zeros for these higher-derivative amplitudes using the method of \cite{Huang:2025blb}, which employs the universal expansions of tree amplitudes. In such expansions, amplitudes from various theories are expressed as a linear combination of bi-adjoint scalar (BAS) amplitudes, where the coefficients are polynomials that depend on the external kinematics \cite{Stieberger:2016lng,Schlotterer:2016cxa,Chiodaroli:2017ngp,Nandan:2016pya,delaCruz:2016gnm,
Fu:2017uzt,Teng:2017tbo,Du:2017kpo,Du:2017gnh,Feng:2019tvb,Zhou:2019gtk,Zhou:2019mbe,Wei:2023yfy,Hu:2023lso,Du:2024dwm,Zhou:2024qwm,Zhou:2024qjh}.
Leveraging the universal expansions of higher-derivative YM and GR amplitudes from \cite{Bonnefoy:2023imz,Zhou:2024qwm} and the well known Kleiss-Kuijf (KK) relation \cite{Kleiss:1988ne}, the hidden zeros for these higher-derivative amplitudes are traced to the hidden zeros for BAS amplitudes, which were rigorously proved in \cite{Zhang:2024efe,Huang:2025blb}.

An important subtlety in the study of hidden zeros within higher-derivative GR amplitudes concerns the emergence of divergences from propagators. The kinematic condition that determines hidden zeros includes, as a key element, the requirement that $k_a\cdot k_b=0$ for specific external particles $a$ and $b$. This relation trivially implies the singularity of the propagator $1/s_{ab}$. For gluon amplitudes, color orderings compatible with the hidden zero condition inherently preclude such divergent propagators. For graviton amplitudes, which are devoid of color ordering, these singularities are inescapable and present a paramount concern for the consistency of the hidden zeros. Fortunately, as we will elucidate, a systematic cancellation of these divergences occurs under the requisite kinematic conditions. This results in an effective expansion of amplitudes that is manifestly finite, thereby placing the proof of hidden zeros on a firm and unambiguous foundation.

The remainder of this paper is structured as follows. In section \ref{sec-expan}, we provide a brief overview of universal expansions for higher-derivative tree amplitudes including $F^3$, $R^2$ and $R^3$. Section \ref{sec-zero-YMandsEFT} applies these expansions to reveal the presence of hidden zeros for higher-derivative YM amplitudes. In section \ref{sec-zero-GR}, we extend this analysis to higher-derivative GR amplitudes, demonstrating their hidden zeros and investigating the cancellations of divergences arising from propagators of the form $1/s_{ab}$. Finally, we conclude with a summary and discussion in section \ref{sec-conclu}.

\section{Universal expansions of higher-derivative tree amplitudes}
\label{sec-expan}

For readers' convenience, in this section we briefly introduce the universal expansions for the higher-derivative tree amplitudes considered in this paper. In subsection \ref{subsec-BASamp}, we rapidly review the BAS amplitudes at the tree-level, which serve as the basis for these expansions. Then, in subsequent subsections, we present the expansions for two types of tree amplitudes: YM amplitudes with a single insertion of the $F^3$ operator, and the GR amplitudes with insertions of the $R^2$ and $R^3$ operators. The descriptions in this section are formal, and the explicit examples of expansions are provided in appendix \ref{sec-example}.

\subsection{The BAS basis}
\label{subsec-BASamp}

The bi-adjoint scalar (BAS) theory describes the cubic interaction of massless scalar field $\phi^{A a}$, with the Lagrangian
\bea
{\cal L}_{\rm BAS}={1\over2}\,\partial_\mu\phi^{Aa}\,\partial^{\mu}\phi^{Aa}+{\lambda\over3!}\,F^{ABC}f^{abc}\,
\phi^{Aa}\phi^{Bb}\phi^{Cc}\,,
\eea
where $F^{ABC}={\rm tr}([T^A,T^B]T^C)$ and $f^{abc}={\rm tr}([T^a,T^b]T^c)$ are usual structure constants of two Lie groups, respectively.

Each tree amplitudes in this theory with coupling constants stripped off consists solely of propagators for massless scalars, and the usual decomposition of group factors leads to
\bea
A^{\rm BAS}_n=\sum_{\sigma\in{\cal S}_n\setminus Z_n}\,\sum_{\sigma'\in{\cal S}'_n\setminus Z'_n}\,
{\rm tr}[T^{A_{\sigma_1}},\cdots T^{A_{\sigma_n}}]\,{\rm tr}[T^{a_{\sigma'_1}}\cdots T^{a_{\sigma'_n}}]\,
{\cal A}^{\rm BAS}_n(\sigma_1,\cdots,\sigma_n|\sigma'_1,\cdots,\sigma'_n)\,,
\eea
where $A_n$ represents the full $n$-point tree amplitude. The summation is over all un-cyclic permutations ${\cal S}_n\setminus Z_n$ and ${\cal S}'_n\setminus Z'_n$. Each partial amplitude ${\cal A}^{\rm BAS}_n(\sigma_1, \cdots, \sigma_n|\sigma'_1, \cdots, \sigma'_n)$ is planar with respect to both orderings $(\sigma_1,\cdots,\sigma_n)$ and $(\sigma'_1,\cdots,\sigma'_n)$. For example, the $5$-point amplitude ${\cal A}^{\rm BAS}_5(1,2,3,4,5|1,2,4,5,3)$ includes only one term, i.e.,
\bea
{\cal A}^{\rm BAS}_5(1,2,3,4,5|1,2,4,5,3)={1\over s_{12}}\,{1\over s_{45}}\,,
\eea
up to an overall $\pm$ sign, since other Feynman diagrams are not compatible with two orderings $(1,2,3,4,5)$ and $(1,2,4,5,3)$ simultaneously.
In the above, each Mandelstam variable $s_{\a}$ is defined as usual
\bea
s_{\a}\equiv k_{\a}^2\,,~~~{\rm with}~k_{\a}\equiv\sum_{\ell\in\a}\,k_\ell\,,~~~~\label{mandelstam}
\eea
where $k_\ell$ is the momentum of the external particle $\ell$, while $\a$ is a subset of external particles $\{1,\cdots,n\}$.
For simplicity, sometimes we denote an $n$-point partial amplitude as ${\cal A}^{\rm BAS}_n({\pmb\sigma}_n|{\pmb\sigma}'_n)$, where ${\pmb\sigma}_n$ and ${\pmb\sigma}'_n$ stand for two orderings.

\begin{figure}
  \centering
  \includegraphics[width=11cm]{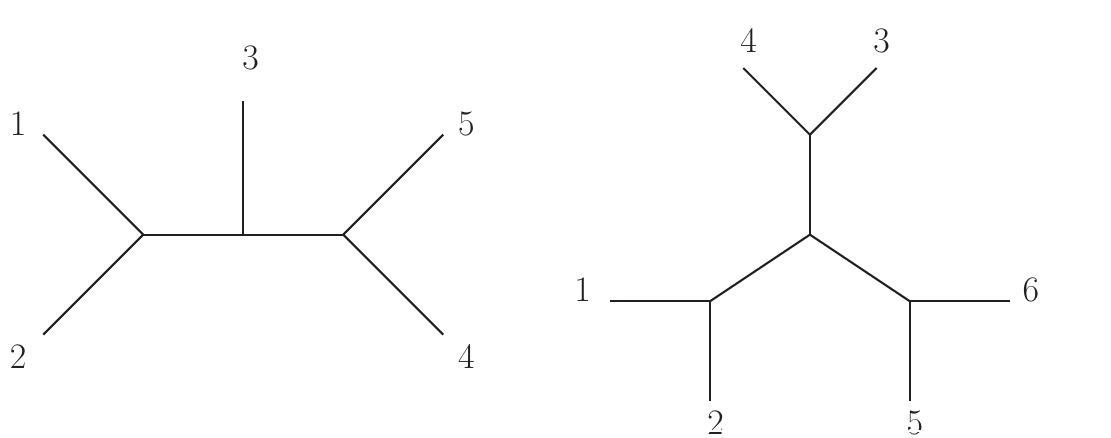} \\
  \caption{Examples of $\Gamma$.}\label{gamma}
\end{figure}

The anti-symmetry of structure constants $F^{ABC}$ and $f^{abc}$ indicates a $-$ sign when swapping two group indices, therefore each partial BAS amplitude always carries
an overall sign $+$ or $-$. This overall sign can be determined by counting flips of external legs, as detailed in \cite{Cachazo:2013iea}. As a consequence of such anti-symmetry, flipping two adjacent legs in one of two orderings creates a relative $-$ sign, that is,
\bea
\Gamma^{\rm BAS}_n(\cdots,p,q,\cdots|\pmb{\sigma}_n)=-\Gamma^{\rm BAS}_n(\cdots,q,p,\cdots|\pmb{\sigma}_n)\,,~~\label{swapp-pq}
\eea
where $\Gamma$, from an individual Feynman diagram, denotes the product of massless scalar propagators excluding the $\pm$ sign, and $\Gamma^{\rm BAS}_n(\pmb\sigma'_n|\pmb\sigma_n)$ is the signed expression that contributes to the amplitude ${\cal A}^{\rm BAS}_n(\pmb\sigma'_n|\pmb\sigma_n)$. As two simple examples, let us consider two diagrams in Fig. \ref{gamma}. For the left diagram, $\Gamma$ is given by
\bea
\Gamma={1\over s_{12}}\,{1\over s_{45}}\,,~~\label{example1-gamma}
\eea
while the corresponding signed expressions are
\bea
&&\Gamma^{\rm BAS}_5(2,1,3,5,4|1,2,3,4,5)={1\over s_{12}}\,{1\over s_{45}}\,,\nn
&&\Gamma^{\rm BAS}_5(1,2,3,5,4|1,2,3,4,5)=-{1\over s_{12}}\,{1\over s_{45}}\,.
\eea
For the right diagram, $\Gamma$ is given as
\bea
\Gamma={1\over s_{12}}\,{1\over s_{34}}\,{1\over s_{56}}\,,~~~~\label{example2-gamma}
\eea
while the signed expressions are
\bea
&&\Gamma^{\rm BAS}_5(2,1,4,3,6,5|1,2,3,4,5,6)={1\over s_{12}}\,{1\over s_{34}}\,{1\over s_{56}}\,,\nn
&&\Gamma^{\rm BAS}_5(2,1,3,4,6,5|1,2,3,4,5,6)=-{1\over s_{12}}\,{1\over s_{34}}\,{1\over s_{56}}\,.
\eea
It is worth to emphasize that the above swapping-relation \eref{swapp-pq} holds for $\Gamma^{\rm BAS}_n(\cdots,p,q,\cdots|\pmb{\sigma}_n)$ and $\Gamma^{\rm BAS}_n(\cdots,q,p,\cdots|\pmb{\sigma}_n)$ from the same $\Gamma$ (it means $\Gamma$ contains the propagator $1/s_{pq}$), but does not hold for full BAS amplitudes in general.
This relation will play the crucial role in section \ref{sec-zero-GR}.

As widely studied, a large variety of tree amplitudes of massless particles can be expanded into double-ordered partial BAS amplitudes \cite{Stieberger:2016lng,Schlotterer:2016cxa,Chiodaroli:2017ngp,Nandan:2016pya,delaCruz:2016gnm,
Fu:2017uzt,Teng:2017tbo,Du:2017kpo,Du:2017gnh,Feng:2019tvb,Zhou:2019gtk,Zhou:2019mbe,Wei:2023yfy,Hu:2023lso,Du:2024dwm,Zhou:2024qwm,Zhou:2024qjh}. The basis of expansions can be chosen via the well known Kleiss-Kuijf (KK) relation \cite{Kleiss:1988ne},
\bea
{\cal A}_n^{\rm BAS}(i,{\pmb\a},j,{\pmb\b}|{\pmb\sigma}_n)=(-)^{|\b|}\,\sum_{\shuffle}\,{\cal A}_n^{\rm BAS}(i,{\pmb\a}\shuffle{\pmb\b}^T,j|{\pmb\sigma}_n)\,,~~~\label{KK}
\eea
which is valid for arbitrary $i,j\in\{1,\cdots,n\}$.
In the above, ${\pmb\b}^T$ stands for the inverse of the ordered set $\pmb\b$. For instance, ${\pmb\b}^T=\{3,2,1\}$ if ${\pmb\b}=\{1,2,3\}$. The summation $\sum_{\shuffle}$ is over all permutations such that the relative order in each of the ordered sets
${\pmb\b}^T$ and $\pmb\a$ is kept. For instance, suppose $\pmb\a=\{a_1,a_2\}$, $\pmb{\b}^T=\{b_1,b_2\}$, then the summation over shuffles
$\shuffle$ reads
\bea
\sum_{\shuffle}\,{\cal A}_6(i,\pmb\a\shuffle{\pmb \b}^T,j)&=&{\cal A}_6(i,a_1,a_2,b_1,b_2,j)+{\cal A}_6(i,a_1,b_1,a_2,b_2,j)+{\cal A}_6(i,a_1,b_1,b_2,a_2,j)\nn
& &+{\cal A}_6(i,b_1,b_2,a_1,a_2,j)+{\cal A}_6(i,b_1,a_1,b_2,a_2,j)+{\cal A}_6(i,b_1,a_1,a_2,b_2,j)\,.~~\label{example-shuffle}
\eea
The KK relation implies that any BAS amplitude can be expanded to BAS amplitudes with two special external legs ($i$ and $j$ in \eref{KK}) fixed at two ends in their left (right) orderings. Hence, it is natural to choose the basis as BAS amplitudes ${\cal A}^{\rm BAS}_n(i,\pmb\sigma_{n-2},j|i,\pmb\sigma'_{n-2},j)$, where $\pmb\sigma_{n-2}$ and $\pmb\sigma'_{n-2}$ are ordered sets for external legs in $\{1,\cdots,n\}\setminus\{i,j\}$. Such choice of basis will be used when expanding unordered amplitudes in subsection \ref{subsec-expan-GRR2R3}. For ordered amplitude ${\cal A}_n(\pmb\sigma_n)$, such like higher-derivative YM amplitudes in subsections \ref{subsec-expan-YMF3}, it is more convenient to choose the basis as ${\cal A}^{\rm BAS}_n(i,\pmb\sigma_{n-2},j|\pmb\sigma_n)$ (or ${\cal A}^{\rm BAS}_n(\pmb\sigma_n|i,\pmb\sigma_{n-2},j)$), where the right (or left) ordering is inherited from ${\cal A}_n(\pmb\sigma_n)$ under consideration. The above bases are called the BAS KK bases.

\subsection{YM amplitudes with single insertion of $F^3$ operator}
\label{subsec-expan-YMF3}

In this subsection, we provide the universal expansion of tree YM amplitudes with a single insertion of local operator $F^3\equiv f^{abc}F^{a~\nu}_{~\mu}F^{b~\rho}_{~\nu}F^{c~\mu}_{~\rho}$. The corresponding Lagrangian is given as
\bea
{\cal L}_{\rm YM}=-{1\over4}\,F^{a}_{\mu\nu}F^{a\mu\nu}-{g\over 3\Lambda^2}\,f^{abc}F^{a~\nu}_{~\mu}F^{b~\rho}_{~\nu}F^{c~\mu}_{~\rho}\,,~~\label{Lag-+2}
\eea
while the Cachazo-He-Yuan (CHY) formula for such amplitudes can be found in \cite{He:2016iqi}.

We can refer to these special higher-derivative ordered YM amplitudes the $F^3$ amplitudes, and denote them as ${\cal A}^{F^3}_n(\pmb\sigma_n)$. As studied in \cite{Bonnefoy:2023imz,Zhou:2024qwm}, each $F^3$ amplitude can be expanded to YM$\oplus$BAS ones as
\bea
{\cal A}_n^{F^3}({\pmb\sigma}_n)=\sum_{\substack{{\pmb{\rho}},\,g\not\in\pmb\rho\\2\leq|\rho|\leq n-1}}\,
{\rm tr}({\rm F}_{{\pmb{\rho}}})\,{\cal A}_n^{\rm YMS}({\pmb{\rho}};{\rm G}_n\setminus\rho|{\pmb{\sigma}}_n)\,.~~\label{YM+2-toYMS}
\eea
In the above, ${\rm G}_n$ represents the full set of external gluons $\{1,\cdots,n\}$, and $\rho$ is a subset of ${\rm G}_n$ without containing the fiducial gluon $g$. The fiducial gluon $g$ can be chosen as any one in ${\rm G}_n$. For a given $\rho$, each ordered set $\pmb\rho$ is created by imposing a specific ordering on its elements. For example, if $\rho=\{1,2\}$, the corresponding $\pmb\rho$ can be $\{1,2\}$ or $\{2,1\}$. The summation is over all possible cyclically inequivalent $\pmb\rho$ with $2\leq|\rho|\leq n-1$, where $|\rho|$ encodes the number of elements in $\rho$.  The ${\cal A}_n^{\rm YMS}({\pmb{\rho}};{\rm G}_n\setminus\rho|{\pmb{\sigma}}_n)$ are tree amplitudes of YM$\oplus$BAS theory, where external legs in $\rho$ are scalars, while those in ${\rm G}_n\setminus\rho$ are gluons. As usual, each BAS scalar enters two orderings which are $\pmb\rho$ and $\pmb\sigma_n$, while each gluon enters only one ordering $\pmb\sigma_n$ among all external legs. For a given $\pmb\rho$, we can label its elements as $\pmb\rho=\{q_1,q_2,\cdots,q_{|\rho|}\}$, the coefficient ${\rm tr}({\rm F}_{\pmb\rho})$ in \eref{YM+2-toYMS} is then expressed as
\bea
{\rm tr}({\rm F}_{\pmb\rho})\equiv\big({\rm F}_{\pmb\rho}\big)_\mu^{~\mu}\,,~~~~{\rm where}~~\big({\rm F}_{\pmb\rho}\big)_\mu^{~\nu}\equiv (-)^{|\rho|} \big(f_{q_1}\big)_\mu^{~\tau_1}(f_{q_2}\big)_{\tau_1}^{~\tau_2}\cdots(f_{q_{|\rho|}}\big)_{\tau_{|\rho|-1}}^{~\nu}\,.~~\label{defin-F}
\eea
Here the anti-symmetric tensor $f_\ell^{\mu\nu}$ is defined as $f_\ell^{\mu\nu}\equiv k_\ell^\mu\epsilon_\ell^\nu-\epsilon_\ell^\mu k_\ell^\nu$, where $k_\ell$ and $\epsilon_\ell$ are momentum and polarization vector of the gluon $\ell$, respectively.

An equivalent alternative expansion formula without the fiducial gluon is \cite{Zhou:2024qwm},
\bea
{\cal A}_n^{F^3}({\pmb\sigma}_n)=\sum_{{\pmb{\rho}},\,2\leq|\rho|\leq n}\,
{\rm tr}({\rm F}_{{\pmb{\rho}}})\,{\cal A}_n^{\rm YMS}({\pmb{\rho}};{\rm G}_n\setminus\rho|{\pmb{\sigma}}_n)\,.~~\label{YM+2-toYMS-alternative}
\eea
In this formula, the summation is over all possible cyclically inequivalent $\pmb\rho$ with $2\leq|\rho|\leq n$. One can also formally extend the length of $\rho$
to be $0\leq|\rho|\leq n$, since the definition of ${\rm tr}({\rm F}_{\pmb\rho})$ ensures ${\rm tr}({\rm F}_{\pmb\rho})=0$ when $|\rho|=0,1$. Although more symmetrical, the second expansion formula \eref{YM+2-toYMS-alternative} includes various redundant terms which cancel each other. For works in this paper, we find the first expansion formula \eref{YM+2-toYMS} to be more effective.

Each YM$\oplus$BAS amplitude can be further expanded to YM$\oplus$BAS amplitudes with more scalars and less gluons \cite{Fu:2017uzt,Teng:2017tbo,Du:2017kpo,Du:2017gnh},
\bea
{\cal A}^{\rm YMS}_n(i,\pmb\sigma_{m-2},j;{\rm G}_{n-m}|\pmb\sigma_n)=\sum_{\pmb\gamma^f}\,\sum_{\shuffle}\,{\rm E}_{\pmb\gamma^f}\,{\cal A}^{\rm YMS}_n(i,\pmb\sigma_{m-2}\shuffle\pmb\gamma^f,j;{\rm G}_{n-m}\setminus\gamma^f|\pmb\sigma_n)\,.~~\label{expan-YMS}
\eea
On the l.h.s, ${\cal A}^{\rm YMS}_n(i,\pmb\sigma_{m-2},j;{\rm G}_{n-m}|\pmb\sigma_n)$ denotes the YM$\oplus$BAS amplitude with $m$ external scalars and $n-m$ external gluons, where $(i,\pmb\sigma_{m-2},j)$ is the ordered set of scalars, while ${\rm G}_{n-m}$ is the unordered set of gluons. On the r.h.s, the summation over shuffles $\shuffle$ is defined around \eref{example-shuffle}. Each $\gamma^f$ is an nonempty subset of ${\rm G}_{n-m}$ which includes the fiducial leg $f$, while each $\pmb\gamma^f$ is obtained by imposing an ordering on elements in $\gamma^f$. The fiducial leg $f$ can be chosen as any element in ${\rm G}_{n-m}$, and should be fixed at the right end of $\pmb\gamma^f$, that is, $\pmb\gamma^f=\{r_1,\cdots,r_{|\gamma-1|},f\}$.
All possible $\pmb\gamma^f$ should be summed.
For a given $\pmb\gamma^f$, the coefficient ${\rm E}_{\pmb\gamma^f}$ in the expansion formula \eref{expan-YMS} is
\bea
{\rm E}_{\pmb\gamma^f}&\equiv&\big(\epsilon_f\big)^{\mu_1} \big(f_{r_{|\gamma|-1}}\big)_{\mu_1}^{~\mu_2}\cdots \big(f_{r_{1}}\big)_{\mu_{|\gamma|-1}}^{~\mu_{|\gamma|}} \big(Y_{r_{1}}\big)_{\mu_{|\gamma|}}\nn
&=&\epsilon_f\cdot f_{r_{|\gamma|-1}}\cdots f_{r_1}\cdot Y_{r_1}\,,~~\label{define-El}
\eea
where the definition of the strength tensor $f_\ell^{\mu\nu}$ was provided after \eref{defin-F}.
In the above, the combinatorial momentum $Y_{r_1}$ is defined as the summation of momenta carried by external legs on the l.h.s of $r_1$
in the ordering $(i,\pmb\sigma_{m-2}\shuffle\pmb\gamma^f,j)$.

By iteratively applying the expansion \eref{expan-YMS}, one can ultimately expand the YM$\oplus$BAS amplitude on the l.h.s of \eref{expan-YMS} into the standard KK basis, with a pair $(i,j)$ of external legs fixed at two ends of the left orderings. On the other hand, one can also perform such recursive expansion to \eref{YM+2-toYMS}, to expand the $F^3$ amplitudes into pure BAS amplitudes. However, BAS amplitudes in the resulting expansion formula do not belong to a special KK basis, since different $\pmb\rho$ in \eref{YM+2-toYMS} has different pair $(i,j)$ at two ends. Nevertheless, one can always choose a particular pair $(i,j)$, and apply the KK relation \eref{KK} to convert all partial BAS amplitudes in the resulting expansion to those in the corresponding KK basis, obtaining
\bea
{\cal A}^{F^3}_n(\pmb\sigma_n)=\sum_{\pmb\a_{n-2}}\,C_{F^3}(\pmb\a_{n-2})\,{\cal A}^{\rm BAS}_n(i,\pmb\a_{n-2},j|\pmb\sigma_n)\,,~~\label{YM+2toKK}
\eea
where $\pmb\a_{n-2}$ encodes ordered sets for external legs in $\{1,\cdots,n\}\setminus\{i,j\}$.
In section \ref{subsec-zero-YMF3}, we will propose an alternative process for expanding $F^3$ amplitudes into the KK basis.

As mentioned earlier, the explicit examples of the expansions in \eref{YM+2-toYMS} and \eref{expan-YMS} can be seen in appendix \ref{sec-example}.

\subsection{$R^2$ and $R^3$ GR amplitudes}
\label{subsec-expan-GRR2R3}

The higher-derivative GR amplitudes under consideration in this paper originate from the low-energy expansion of bosonic closed string theory,
\bea
S=-{2\over\kappa^2}\,\int\,d^4x\sqrt{-g}\,\Big[R-2\,(\partial_\mu\phi)^2-{1\over12}\,H^2+{\a'\over4}\,e^{-2\phi}\,G_2+\a'^2\,e^{-4\phi}\,
\big({I_1\over48}+{G_3\over24}\big)+{\cal O}(\a'^3)\Big]\,,~~\label{lowenergy-string}
\eea
where $G_2$ is known as the conventional Gauss-Bonnet term containing two powers of Riemann tensors, while $I_1$ and $G_3$ contain three powers of Riemann tensors. When the external states are restricted to pure gravitons, each tree amplitude at the $\alpha'$ order receives contributions solely from a single insertion of the $G_2$ operator. At the $\alpha'^2$ order, the amplitude includes contributions from both a single insertion of the $I_1$ or $G_3$ operator, and double insertions of $R^2$ operators with the exchange of an intermediate dilaton. The pure graviton amplitudes at the $\alpha'$ and $\alpha'^2$ orders introduced above, are referred to as $R^2$ and $R^3$ amplitudes, respectively.

The Cachazo-He-Yuan (CHY) integrands for $n$-point tree YM, $F^3$, $R^2$ and $R^3$ amplitudes, are presented as \cite{Cachazo:2013hca,Cachazo:2013iea,He:2016iqi}
\bea
& &{\cal I}_n^{{\rm YM}}({\pmb\sigma}_n)={\rm Pf}'{\Psi}(\epsilon)\,{\rm PT}({\pmb\sigma}_n)\,,~~~~{\cal I}_n^{F^3}({\pmb\sigma}_n)={\cal P}_n(\epsilon)\,{\rm PT}({\pmb\sigma}_n)\,,\nn
& &{\cal I}_n^{R^2}={\cal P}_n(\epsilon)\,{\rm Pf}'{\Psi}(\W\epsilon)\,,
~~~~~~~~~~~~~~{\cal I}_n^{R^3}={\cal P}_n(\epsilon)\,{\cal P}_n(\W\epsilon)\,,~~\label{CHY-integrand}
\eea
where ${\rm PT}({\pmb\sigma}_n)$ is the so called Parke-Taylor factor without containing any kinematic variable, ${\rm Pf}'{\Psi}(\epsilon)$ and ${\cal P}_n(\epsilon)$ (or ${\rm Pf}'{\Psi}(\W\epsilon)$ and ${\cal P}_n(\W\epsilon)$) are polynomials of Lorentz invariants, depending on external momenta $\{k_\ell\}$ and polarization vectors $\{\epsilon_\ell\}$ (or $\{\W\epsilon_\ell\}$). The polarization tensor of a graviton is decomposed as $\epsilon_\ell^{\mu\nu}=\epsilon_\ell^\mu\W\epsilon_\ell^\nu$,
where $\W\epsilon_\ell^\mu=\epsilon_\ell^\mu$. The integrand ${\cal I}^{F^3}_n$ in \eref{CHY-integrand} implies that the expansion in \eref{YM+2toKK} can be understood as
\bea
{\cal P}_n(\epsilon)=\sum_{{\pmb\a}_{n-2}}\,C_{F^3}(\epsilon,{\pmb\a}_{n-2})\,{\rm PT}(i,{\pmb\a}_{n-2},j)\,,~~\label{expan-P}
\eea
since the CHY integrand of each BAS amplitude ${\cal A}^{\rm BAS}_n(\pmb\sigma_n|\pmb\sigma'_n)$ is
\bea
{\cal I}_n^{\rm BAS}({\pmb\sigma}_n|{\pmb\sigma}'_n)={\rm PT}({\pmb\sigma}_n)\,{\rm PT}({\pmb\sigma}'_n)\,.~~\label{CHY-BAS}
\eea
In the above, we relabeled coefficients $C_{F^3}({\pmb\a}_{n-2})$ in \eref{YM+2toKK} as $C_{F^3}(\epsilon,{\pmb\a}_{n-2})$, to distinguish two sets of polarizations $\{\epsilon_\ell\}$ and $\{\W\epsilon_\ell\}$. Similarly, the polynomial ${\rm Pf}'{\Psi}(\epsilon)$ can be expanded as
\bea
{\rm Pf}'{\Psi}(\epsilon)=\sum_{{\pmb\a}_{n-2}}\,C_{\rm YM}(\epsilon,{\pmb\a}_{n-2})\,{\rm PT}(i,{\pmb\a}_{n-2},j)\,.~~\label{expan-pfaffian}
\eea
This expansion of integrands leads to the following expansion of amplitudes
\bea
{\cal A}^{\rm YM}_n(\pmb\sigma_n)=\sum_{\pmb\a_{n-2}}\,C_{\rm YM}(\epsilon,\pmb\a_{n-2})\,{\cal A}^{\rm BAS}_n(i,\pmb\a_{n-2},j|\pmb\sigma_n)\,.~~\label{expan-YM}
\eea

Let us rapidly introduce the construction for coefficients $C_{\rm YM}(\epsilon,\pmb\a_{n-2})$ \cite{Fu:2017uzt,Teng:2017tbo,Du:2017kpo,Du:2017gnh}.
To achieve the expansion formula \eref{expan-YM}, the central step is to expand the YM amplitudes into YM$\oplus$BAS ones as
\bea
{\cal A}^{\rm YM}_n(\pmb\sigma_n)=\sum_{\pmb\pi,0\leq|\pi|\leq n-2}\,{\rm T}^{(i,j)}_{\pmb\pi}\,{\cal A}^{\rm YMS}_n(i,\pmb\pi,j;{\rm G}_n\setminus\{\{i,j\}\cup\pi\}|\pmb\sigma_n)\,,~~\label{YMtoYMS}
\eea
where the summation is over all ordered sets $\pmb\pi$ satisfying $\pi\subset{\rm G}_n\setminus\{i,j\}$, and $0\leq|\pi|\leq n-2$. Each ordered set $\pmb\pi=\{p_1,\cdots,p_{|\pi|}\}$ corresponds to a kinematic factor ${\rm T}^{(i,j)}_{\pmb\pi}$, defined as
\bea
{\rm T}^{(i,j)}_{\pmb\pi}&\equiv& (-)^{|\pi|}\,\big(\epsilon_i\big)^{\mu_1}\big(f_{p_1}\big)_{\mu_1}^{~\mu_2}\cdots\big(f_{p_{|\pi|}}\big)_{\mu_{|\pi|}}^{~\mu_{|\pi|+1}}
\big(\epsilon_j\big)_{\mu_{|\pi|+1}}\nn
&=&(-)^{|\pi|}\,\epsilon_i\cdot f_{p_1}\cdots f_{p_{|\pi|}}\cdot \epsilon_j\,,~~\label{define-T}
\eea
The YM$\oplus$BAS amplitudes in \eref{YMtoYMS} can be further expanded into BAS amplitudes in the KK basis, by iteratively applying the recursive expansion \eref{expan-YMS}. After finishing the above manipulation, the pure YM amplitude is ultimately expanded into the KK basis with $(i,j)$ fixed at two ends of the left orderings. The corresponding coefficients $C_{\rm YM}(\epsilon,\pmb\a_{n-2})$ in \eref{expan-YM} are then obtained.

Substituting expansions \eref{expan-P} and \eref{expan-pfaffian} into CHY integrands of $R^2$ and $R^3$ in \eref{CHY-integrand} gives rise to the expansions of $R^2$ and $R^3$ amplitudes, expressed as
\bea
{\cal A}^{R^2}_n=\sum_{\pmb\a_{n-2}}\,\sum_{\pmb\a'_{n-2}}\,C_{F^3}(\epsilon,\pmb\a_{n-2})\,C_{{\rm YM}}(\W\epsilon,\pmb\a'_{n-2})\,
{\cal A}^{\rm BAS}_n(i,\pmb\a_{n-2},j|i,\pmb\a'_{n-2},j)\,,~~\label{expan-R2}
\eea
and
\bea
{\cal A}^{R^3}_n=\sum_{\pmb\a_{n-2}}\,\sum_{\pmb\a'_{n-2}}\,C_{F^3}(\epsilon,\pmb\a_{n-2})\,C_{F^3}(\W\epsilon,\pmb\a'_{n-2})\,
{\cal A}^{\rm BAS}_n(i,\pmb\a_{n-2},j|i,\pmb\a'_{n-2},j)\,.~~\label{expan-R3}
\eea
%

\section{Hidden zeros for higher-derivative YM amplitudes}
\label{sec-zero-YMandsEFT}

In this section, we employ the universal expansion of $F^3$ tree amplitudes (see section \ref{subsec-expan-YMF3}) to establish the existence of their hidden zeros. As our approach traces these zeros back to those in bi-adjoint scalar (BAS) amplitudes, we begin with a brief review of hidden zeros for BAS amplitudes in subsection \ref{subsec-0-BAS}. We then present the corresponding hidden zeros for $F^3$ amplitudes in subsection \ref{subsec-zero-YMF3}, where we also explain the core reason for this phenomenon from the perspective of the universal expansion---namely, the reduced expansion \eref{expan-YM+2-reduced} and the related discussion. The supporting examples and a general proof are provided in subsections \ref{subsec-example-4ptYM+2} through \ref{subsec-proof-YM+2}.

\subsection{Hidden zeros for BAS amplitudes}
\label{subsec-0-BAS}

A special set of double ordered BAS amplitudes vanish on certain loci in kinematic space. Such hidden zeros can be achieved as follows. For $n$-point BAS amplitudes, one can choose two external legs $\hat{i}$ and $\hat{j}$, and divide the remaining external legs into two subsets $A$ and $B$, namely $A\cup B=\{1,\cdots,n\}\setminus\{\hat{i},\hat{j}\}$.
The amplitudes ${\cal A}^{\rm BAS}_n(\pmb\sigma_n|\pmb\sigma'_n)$ with two orderings satisfying
\bea
{\pmb\sigma}_n=\pmb A,\hat{i},\pmb B,\hat{j}\,,~~~~{\pmb\sigma}'_n=\pmb A',\hat{i},\pmb B',\hat{j}\,,~~~~{\rm up~to~cyclic~permutations}\,,~~\label{compa-order}
\eea
vanish as
\bea
{\cal A}^{\rm BAS}_n(\pmb A,\hat{i},\pmb B,\hat{j}|\pmb A',\hat{i},\pmb B',\hat{j})&\xrightarrow[]{\eref{condi0-momentum}}&0\,,~~\label{zero-BAS}
\eea
where the kinematic condition is given as
\bea
k_a\cdot k_b=0\,,~~~~{\rm for}~~\forall\,a\in A\,,~~\forall\,b\in B\,.~~\label{condi0-momentum}
\eea
In \eref{compa-order}, $\pmb A$ and $\pmb A'$ are two ordered sets obtained by imposing orderings on elements in $A$, while $\pmb B$ and $\pmb B'$ are ordered sets obtained by imposing orderings on elements in $B$. The hidden zeros in \eref{zero-BAS} can be proved via either the CHY formula or Feynman rules \cite{Zhang:2024efe,Huang:2025blb}.

It is worth emphasizing that not all BAS amplitudes exhibit hidden zeros. For a given choice of $A$ and $B$, the hidden zeros exist in BAS amplitudes with two orderings compatible with the choice of $A$ and $B$. That is, in either $\pmb\sigma_n$ or $\pmb\sigma'_n$, elements in $A$ and $B$ are separated by $\hat{i}$ and $\hat{j}$ as in \eref{compa-order}.

As the simplest examples, let us consider two $4$-point BAS amplitudes. We choose $\hat{i}$, $\hat{j}$, $A$ and $B$ as
\bea
\hat{i}=1\,,~~\hat{j}=3\,,~~A=\{2\}\,,~~B=\{4\}\,.~~\label{choice-4pt}
\eea
The first example is the amplitude ${\cal A}^{\rm BAS}_4(1,2,3,4|1,4,3,2)$. In this example, both two orderings are compatible with the choice \eref{choice-4pt}, therefore the amplitude exhibits hidden zero. Such hidden zero can be verified by directly calculating this amplitude,
\bea
{\cal A}^{\rm BAS}_4(1,2,3,4|1,4,3,2)={1\over s_{12}}+{1\over s_{14}}=-{s_{13}\over s_{12}s_{14}}\,,
\eea
which vanishes when $s_{13}=2k_2\cdot k_4=0$. The second example is the amplitude ${\cal A}^{\rm BAS}_4(1,2,3,4|1,3,2,4)$. It does not exhibit hidden zero, since the right ordering $(1,3,2,4)$ is not compatible with the choice \eref{choice-4pt}. As can be seen, the explicit formula
\bea
{\cal A}^{\rm BAS}_4(1,2,3,4|1,3,2,4)=-{1\over s_{14}}
\eea
is manifestly nonzero when $k_2\cdot k_4=0$.

\subsection{Hidden zeros for $F^3$ amplitudes}
\label{subsec-zero-YMF3}

As we will prove, the $F^3$ amplitudes has the following hidden zeros
\bea
{\cal A}^{F^3}_n(\hat{i},\pmb A,\hat{j},\pmb B)&\xrightarrow[]{\eref{condi0-YM}}&0\,,~~\label{zero-YM+2}
\eea
on the special loci in kinematic space determined by
\bea
k_a\cdot k_b=k_a\cdot\epsilon_b=\epsilon_a\cdot k_b=\epsilon_a\cdot\epsilon_b=0\,.~~\label{condi0-YM}
\eea
The above kinematic condition is obtained by extending \eref{condi0-momentum} to include polarization vectors. The ordering of amplitude is again compatible with the choice of $A$ and $B$, i.e., elements in $A$ and $B$ are separated by $\hat{i}$ and $\hat{j}$.

To prove the behavior \eref{zero-YM+2}, we will utilize the universal expansion of $F^3$ amplitudes described in section \ref{subsec-expan-YMF3}, as well as the hidden zeros for BAS amplitudes in \eref{zero-BAS}. In general, the BAS amplitudes appear in the universal expansion do not carry two orderings compatible with the special choice of $A$ and $B$ in \eref{condi0-momentum}, thus the hidden zeros for BAS amplitudes cannot be applied directly. Nevertheless, as will be seen, the kinematic condition \eref{condi0-YM} allows us to convert all BAS amplitudes in the expansion to those in \eref{zero-BAS} with compatible orderings, through the KK relation in \eref{KK}. More explicitly, on the specific loci \eref{condi0-YM}, the expansion of $F^3$ amplitudes can be reduced to
\bea
{\cal A}_n^{F^3}(\hat{i},\pmb A,\hat{j},\pmb B)&\xrightarrow[]{\eref{condi0-YM}}&\sum_{\pmb A'}\,\sum_{\pmb B'}\,C_{F^3}(\epsilon,\pmb A',\pmb B')\,\sum_{\shuffle}\,{\cal A}^{\rm BAS}_n(\hat{i},\pmb A'\shuffle\pmb B',\hat{j}|\hat{i},\pmb A,\hat{j},\pmb B)\,,~~\label{expan-YM+2-reduced}
\eea
where $\pmb A'$ and $\pmb B'$ are ordered amplitudes obtained by imposing orderings on elements in $A$ and $B$, respectively. Since each $C(\epsilon,\pmb A',\pmb B')$ is independent of the shuffles $\shuffle$, the KK relation \eref{KK} can be applied, and all ${\cal A}^{\rm BAS}_n(\hat{i},\pmb A'\shuffle\pmb B',\hat{j}|\hat{i},\pmb A,\hat{j},\pmb B)$ are turned to ${\cal A}^{\rm BAS}_n(\hat{i},\pmb A',\hat{j},\pmb B'^T|\hat{i},\pmb A,\hat{j},\pmb B)$, satisfying the hidden zero condition. The $F^3$ amplitudes then vanish as in \eref{zero-YM+2}, follows from the vanishing of BAS amplitudes in \eref{zero-BAS}.

The above discussion shows that the reduced expansion \eref{expan-YM+2-reduced} plays the central role in our proof for the hidden zeros for $F^3$ amplitudes.
In subsection \ref{subsec-example-4ptYM+2} and subsection \ref{subsec-example-5ptYM+2}, we will give two examples to illustrate the emergence of the reduced expansion \eref{expan-YM+2-reduced}. The general proof for such reduced expansion will be presented in subsection \ref{subsec-proof-YM+2}.

\subsection{Example: $4$-point amplitude}
\label{subsec-example-4ptYM+2}

In this subsection, we consider the simplest example, the hidden zero for the $4$-point amplitude ${\cal A}^{F^3}_4(1,2,3,4)$.
We choose $\hat{i}=1$, $\hat{j}=3$, $A=\{2\}$, $B=\{4\}$, the zero kinematics \eref{condi0-YM} is then simplified to
\bea
k_2\cdot k_4=k_2\cdot\epsilon_4=\epsilon_2\cdot k_4=\epsilon_2\cdot\epsilon_4=0\,.~~\label{condi-YM-4pt}
\eea
Because of this kinematic condition, the expansion \eref{YM+2-toYMS} is reduced to
\bea
{\cal A}^{F^3}_4(1,2,3,4)&\xrightarrow[]{\eref{condi-YM-4pt}}&{\rm tr}(F_{23})\,{\cal A}^{\rm YMS}_4(2,3;\underline{1,4}|1,2,3,4)\,+\,{\rm tr}(F_{43})\,{\cal A}^{\rm YMS}_4(4,3;\underline{1,2}|1,2,3,4)\,,~~\label{4ptYM-step1}
\eea
where the fiducial leg is chosen to be $g=\hat{i}=1$. In the above, the effective cyclically inequivalent $\pmb\rho$ are found to be $\pmb\rho=\{2,3\}$ and $\pmb\rho=\{4,3\}$, since \eref{condi-YM-4pt} forces ${\rm tr}(F_{\pmb\rho})=0$ when $\pmb\rho$ contains $2$ and $4$ simultaneously. In the above, we used $\underline{1,4}$ and $\underline{1,2}$ to label unordered sets $\{1,4\}$ and $\{1,2\}$.

One can iteratively apply the expansion \eref{expan-YMS} to expand YMS$\oplus$BAS amplitudes in \eref{4ptYM-step1} into BAS KK basis, with $\hat{i}=1$ and $\hat{j}=3$ fixed at two ends in the left orderings. That is, we choose the pair $(i,j)$ for the KK basis as $(i,j)=(\hat{i},\hat{j})$. Let us focus on the first term ${\rm tr}(F_{23})\,{\cal A}^{\rm YMS}_4(2,3;\underline{1,4}|1,2,3,4)$ on the r.h.s of \eref{4ptYM-step1}. We choose the fiducial leg in \eref{expan-YMS} as $f=\hat{i}=1$, then the allowed $\pmb\gamma^1$ has only one candidate $\pmb\gamma^1=\{1\}$, since the kinematic factor ${\rm E}_{\pmb\gamma^1}$ defined in \eref{define-El} vanishes when $\pmb\gamma^1=\{4,1\}$,
due to the kinematic condition \eref{condi-YM-4pt} as well as the observation $Y_4=k_2$. Thus we arrive at
\bea
{\rm tr}(F_{23})\,{\cal A}^{\rm YMS}_4(2,3;\underline{1,4}|1,2,3,4)&\xrightarrow[]{\eref{condi-YM-4pt}}&{\rm tr}(F_{23})\,(\epsilon_1\cdot k_2)\,{\cal A}^{\rm YMS}_4(2,1,3;\underline{4}|1,2,3,4)\,.~~\label{4ptYM-step2}
\eea

Since our purpose is to expand the amplitude into the BAS KK basis with $(1,3)$ fixed at two ends of the left orderings, it is convenient to turn the left ordering $(2,1,3)$ in ${\cal A}^{\rm YMS}_4(2,1,3;\underline{4}|1,2,3,4)$ to another one $(1,2,3)$. Such transmutation is based on the fact that the KK relation is also valid for purely scalars in the YM$\oplus$BAS amplitude ${\cal A}^{\rm YMS}_n(i,\pmb\a,j,\pmb\b;{\rm G}_{n-m}|\pmb\sigma_n)$, that is,
\bea
{\cal A}^{\rm YMS}_n(i,\pmb\a,j,\pmb\b;{\rm G}_{n-m}|\pmb\sigma_n)=(-)^{|\b|}\,\sum_{\shuffle}\,{\cal A}^{\rm YMS}_n(i,\pmb\a\shuffle\pmb\b^T,j;{\rm G}_{n-m}|\pmb\sigma_n)\,.~~\label{KK-generalized}
\eea
The above generalized version of KK relation can be proved via various methods. For instance, as shown in \cite{Feng:2019tvb}, it can be proved by utilizing differential operators proposed in \cite{Cheung:2017ems} which transmute pure YM amplitudes to YM$\oplus$BAS ones. Using \eref{KK-generalized}, one can convert the r.h.s of \eref{4ptYM-step2} to
\bea
{\rm tr}(F_{23})\,(\epsilon_1\cdot k_2)\,{\cal A}^{\rm YMS}_4(2,1,3;\underline{4}|1,2,3,4)=-{\rm tr}(F_{23})\,(\epsilon_1\cdot k_2)\,{\cal A}^{\rm YMS}_4(1,2,3;\underline{4}|1,2,3,4)\,.~~\label{4ptYM-step3}
\eea
Then, we apply the expansion \eref{expan-YMS} once again, to turn the YM$\oplus$BAS amplitude on the r.h.s of \eref{4ptYM-step3} to the pure BAS amplitudes as
\bea
&&{\rm tr}(F_{23})\,(\epsilon_1\cdot k_2)\,{\cal A}^{\rm YMS}_4(1,2,3;\underline{4}|1,2,3,4)\xrightarrow[]{\eref{condi-YM-4pt}}\nn
&&~~~~~~~~~~~~~~~~~~~~~~~~{\rm tr}(F_{23})\,(\epsilon_1\cdot k_2)\,(\epsilon_4\cdot k_1)\,\sum_{\shuffle}\,{\cal A}^{\rm BAS}_4(1,2\shuffle 4,3|1,2,3,4)\,.
\eea
Here the key observation is, the effective part of $Y_4$ is always $Y^{\rm eff}_4=k_1$ for both two orderings $(1,2,4,3)$ and $(1,4,2,3)$, according to the kinematic condition $\epsilon_4\cdot k_2=0$ in \eref{condi-YM-4pt}.
Thus, the l.h.s of \eref{4ptYM-step2} is expanded as
\bea
{\rm tr}(F_{23})\,{\cal A}^{\rm YMS}_4(2,3;\underline{1,4}|1,2,3,4)&\xrightarrow[]{\eref{condi-YM-4pt}}&-{\rm tr}(F_{23})\,(\epsilon_1\cdot k_2)\,(\epsilon_4\cdot k_1)\,\sum_{\shuffle}\,{\cal A}^{\rm BAS}_4(1,2\shuffle 4,3|1,2,3,4)\,.~~\label{4ptYM-step4}
\eea

The treatment for the second term on the r.h.s of \eref{4ptYM-step1} is analogous. One can repeat the above process to obtain
\bea
{\rm tr}(F_{43})\,{\cal A}^{\rm YMS}_4(4,3;\underline{1,2}|1,2,3,4)=-{\rm tr}(F_{43})\,(\epsilon_1\cdot k_4)\,(\epsilon_2\cdot k_1)\,{\cal A}^{\rm BAS}_4(1,2\shuffle 4,3|1,2,3,4)\,.~~\label{4ptYM-part2}
\eea
Combining \eref{4ptYM-step4} and \eref{4ptYM-part2} together, we get
\bea
&&{\cal A}^{F^3}_4(1,2,3,4)\,\xrightarrow[]{\eref{condi-YM-4pt}}\nn
&&-\Big[{\rm tr}(F_{23})\,(\epsilon_1\cdot k_2)\,(\epsilon_4\cdot k_1)+{\rm tr}(F_{43})\,(\epsilon_1\cdot k_4)\,(\epsilon_2\cdot k_1)\Big]\,\sum_{\shuffle}\,{\cal A}^{\rm BAS}_4(1,2\shuffle 4,3|1,2,3,4)\,,
\eea
satisfying the reduced expansion in \eref{expan-YM+2-reduced}. The KK relation \eref{KK} converts the BAS amplitudes in the above expansion as
\bea
\sum_{\shuffle}\,{\cal A}^{\rm BAS}_4(1,2\shuffle 4,3|1,2,3,4)=-{\cal A}^{\rm BAS}_4(1,2,3,4|1,2,3,4)\,,
\eea
thus the hidden zero for ${\cal A}^{F^3}_4(1,2,3,4)$ follows from the zero for BAS amplitude ${\cal A}^{\rm BAS}_4(1,2,3,4|1,2,3,4)$ in \eref{zero-BAS}, i.e.,
\bea
{\cal A}^{\rm BAS}_4(1,2,3,4|1,2,3,4)&\xrightarrow[]{k_2\cdot k_4=0}&0\,.
\eea
%

\subsection{Example: $5$-point amplitude}
\label{subsec-example-5ptYM+2}

In this subsection we consider the $5$-point example ${\cal A}^{F^3}_5(1,2,3,4,5)$. This example is complicated enough to illustrate most of mechanisms for the emergence of reduced expansion \eref{expan-YM+2-reduced}. We will omit various details which bear strong similarity with those in the previous subsection \ref{subsec-example-4ptYM+2}, and focus on new situations.

Let us choose $\hat{i}=1$, $\hat{j}=4$, $A=\{2,3\}$, $B=\{5\}$, the kinematic condition \eref{condi0-YM} then reads
\bea
\{k_2,k_3\}\cdot k_5=\{k_2,k_3\}\cdot\epsilon_5=\{\epsilon_2,\epsilon_3\}\cdot k_5=\{\epsilon_2,\epsilon_3\}\cdot\epsilon_5=0\,.~~\label{condi-YM-5pt}
\eea
The expansion \eref{YM+2-toYMS} is reduced to
\bea
{\cal A}^{F^3}_5(1,2,3,4,5)\,\xrightarrow[]{\eref{condi-YM-5pt}}&&{\rm tr}({\rm F}_{23})\,{\cal A}^{\rm YMS}_5(2,3;\underline{1,4,5}|1,2,3,4,5)+{\rm tr}({\rm F}_{24})\,{\cal A}^{\rm YMS}_5(2,4;\underline{1,3,5}|1,2,3,4,5)\nn
+&&{\rm tr}({\rm F}_{34})\,{\cal A}^{\rm YMS}_5(3,4;\underline{1,2,5}|1,2,3,4,5)+{\rm tr}({\rm F}_{54})\,{\cal A}^{\rm YMS}_5(5,4;\underline{1,2,3}|1,2,3,4,5)\nn
+&&{\rm tr}({\rm F}_{234})\,{\cal A}^{\rm BAS}_5(2,3,4;\underline{1,5}|1,2,3,4,5)\nn
+&&{\rm tr}({\rm F}_{324})\,{\cal A}^{\rm BAS}_5(3,2,4;\underline{1,5}|1,2,3,4,5)\,,~~\label{5ptYM-step1}
\eea
since other cyclically inequivalent $\pmb\rho$---which contains elements from $A$ and $B$ simultaneously---correspond to ${\rm tr}({\rm F}_{\pmb\rho})=0$,
as implied by the kinematic condition \eref{condi-YM-5pt}.

In the first term on the r.h.s of \eref{5ptYM-step1}, elements in $\pmb\rho_1=\{2,3\}$ are solely from $A$. This is a new situation which does not happen in the $4$-point example, thus we will show the detailed treatment for this term. We choose the fiducial legs in \eref{expan-YMS} as $f=\hat{j}=4$, and expand this term into three parts as
\bea
{\rm tr}({\rm F}_{23})\,{\cal A}^{\rm YMS}_5(2,3;\underline{1,4,5}|1,2,3,4,5)&\xrightarrow[]{\eref{condi-YM-5pt}}&{\rm tr}({\rm F}_{23})\,\Big[(\epsilon_4\cdot k_2)\,{\cal A}^{\rm YMS}_5(2,4,3;\underline{1,5}|1,2,3,4,5)\nn
&&+(\epsilon_4\cdot f_1\cdot k_2)\,{\cal A}^{\rm YMS}_5(2,1,4,3;\underline{5}|1,2,3,4,5)\nn
&&+(\epsilon_4\cdot f_5\cdot f_1\cdot k_2)\,{\cal A}^{\rm BAS}_5(2,1,5,4,3|1,2,3,4,5)\Big]\,,~~\label{5pt-term1}
\eea
with $\pmb\gamma^4=\{4\}$, $\pmb\gamma^4=\{1,4\}$ and $\pmb\gamma^4=\{1,5,4\}$, respectively. For the first term on the r.h.s of \eref{5pt-term1}, we use \eref{expan-YMS} to expand it as
\bea
&&{\rm tr}({\rm F}_{23})\,(\epsilon_4\cdot k_2)\,{\cal A}^{\rm YMS}_5(2,4,3;\underline{1,5}|1,2,3,4,5)\nn
=&&{\rm tr}({\rm F}_{23})\,(\epsilon_4\cdot k_2)\,{\cal A}^{\rm YMS}_5(3,2,4;\underline{1,5}|1,2,3,4,5)\nn
\xrightarrow[]{\eref{condi-YM-5pt}}&&{\rm tr}({\rm F}_{23})\,(\epsilon_4\cdot k_2)\,(\epsilon_1\cdot k_3)\,{\cal A}^{\rm YMS}_5(3,1,2,4;\underline{5}|1,2,3,4,5)\nn
&&+{\rm tr}({\rm F}_{23})\,(\epsilon_4\cdot k_2)\,(\epsilon_1\cdot k_{32})\,{\cal A}^{\rm YMS}_5(3,2,1,4;\underline{5}|1,2,3,4,5)\nn
=&&-{\rm tr}({\rm F}_{23})\,(\epsilon_4\cdot k_2)\,(\epsilon_1\cdot k_3)\,\sum_{\shuffle'}\,{\cal A}^{\rm YMS}_5(1,2\shuffle'3,4;\underline{5}|1,2,3,4,5)\nn
&&+{\rm tr}({\rm F}_{23})\,(\epsilon_4\cdot k_2)\,(\epsilon_1\cdot k_{32})\,{\cal A}^{\rm YMS}_5(1,2,3,4;\underline{5}|1,2,3,4,5)\nn
\xrightarrow[]{\eref{condi-YM-5pt}}&&-{\rm tr}({\rm F}_{23})\,(\epsilon_4\cdot k_2)\,(\epsilon_1\cdot k_3)\,(\epsilon_5\cdot k_1)\,\sum_{\shuffle'}\,\sum_{\shuffle}\,{\cal A}^{\rm BAS}_5(1,2\shuffle'3\shuffle5,4|1,2,3,4,5)\nn
&&+{\rm tr}({\rm F}_{23})\,(\epsilon_4\cdot k_2)\,(\epsilon_1\cdot k_{23})\,(\epsilon_5\cdot k_1)\,\sum_{\shuffle}\,{\cal A}^{\rm BAS}_5(1,\{2,3\}\shuffle5,4|1,2,3,4,5)\,.~~\label{5pt-term1-part1}
\eea
In the first step of \eref{5pt-term1-part1}, we have used the cyclic equivalence between two orderings $(2,4,3)$ and $(3,2,4)$. In the second step, the fiducial leg is chosen to be $f=\hat{j}=1$. At this step, the possibility $\pmb\gamma^1=\{5,1\}$ is excluded by observing $Y_5=k_3\,{\rm or}\,k_{32}$, which implies ${\rm E}_{51}=\epsilon_1\cdot f_5\cdot Y_5=0$ due to the kinematic condition \eref{condi-YM-5pt}. The third step uses the generalized KK relation \eref{KK-generalized} to transmute YM$\oplus$BAS amplitudes to those with $(1,4)$ fixed at two ends in left orderings. The last step converts the final gluon $5$ to the scalar particle. In the last step, the key observation is that the effective part of $Y_5$ is $Y^{\rm eff}_5=k_1$ for any $\shuffle$.

Comparing the final form in \eref{5pt-term1-part1} with the general reduced expansion \eref{expan-YM+2-reduced}, we see that \eref{5pt-term1-part1} exhibits the basic character of \eref{expan-YM+2-reduced}. The ordered sets $\pmb A'$ are given by $\{2,3\}$ or $\{3,2\}$, while $\pmb B'=\{5\}$. Two candidates of $\pmb A'$ are summed, with appropriate coefficients which are independent of the shuffles labeled as $\shuffle$.

The second term on the r.h.s of \eref{5pt-term1} can be further expanded as
\bea
&&{\rm tr}({\rm F}_{23})\,(\epsilon_4\cdot f_1\cdot k_2)\,{\cal A}^{\rm YMS}_5(2,1,4,3;\underline{5}|1,2,3,4,5)\,\xrightarrow[]{\eref{condi-YM-5pt}}\nn
&&~~~~~~~~~~~~~~~~{\rm tr}({\rm F}_{23})\,(\epsilon_4\cdot f_1\cdot k_2)\,(\epsilon_5\cdot k_1)\,\sum_{\shuffle}\,{\cal A}^{\rm BAS}_5(1,\{2,3\}\shuffle5,4|1,2,3,4,5)\,.
\eea
The details are omitted, due to the similarity. For the third term in \eref{5pt-term1}, an important observation is that elements from $A$ and $B$ are already separated by $\hat{i}$ and $\hat{j}$ in the left ordering $(2,1,5,4,3)$. The vanishing of this BAS amplitude is automatic, since both two orderings are compatible with the hidden zero condition of BAS.

The remaining terms in \eref{5ptYM-step1} correspond to
\bea
&&\pmb\rho_2=\{2,4\}\,,~~~~\pmb\rho_3=\{3,4\}\,,~~~~\pmb\rho_4=\{5,4\}\,,\nn
&&\pmb\rho_5=\{2,3,4\}\,,~~~~~~\pmb\rho_6=\{3,2,4\}\,.
\eea
These ordered sets $\pmb\rho$ have a common feature---one element in each of them is $\hat{j}=4$, while other elements are solely from $A$ or $B$.
We have encountered such situation in the previous $4$-point example, and the treatment for these terms is extremely similar.
By applying the technic in the previous subsection \ref{subsec-example-4ptYM+2}, one can show that each of them can be expanded as in the reduced expansion formula \eref{expan-YM+2-reduced}. For instances,
\bea
&&{\rm tr}({\rm F}_{24})\,{\cal A}^{\rm YMS}_5(2,4;\underline{1,3,5}|1,2,3,4,5)\,\xrightarrow[]{\eref{condi-YM-5pt}}\nn
&&~~~~~~~~~~~~~~~~{\rm tr}({\rm F}_{24})\,\Big[-(\epsilon_1\cdot k_2)\,(\epsilon_3\cdot k_1)\,(\epsilon_5\cdot k_1)\,\sum_{\shuffle}\,{\cal A}^{\rm BAS}_5(1,\{3,2\}\shuffle5,4|1,2,3,4,5)\nn
&&~~~~~~~~~~~~~~~~+(\epsilon_1\cdot f_3\cdot k_2)\,(\epsilon_5\cdot k_1)\,\sum_{\shuffle}\,{\cal A}^{\rm BAS}_5(1,\{3,2\}\shuffle5,4|1,2,3,4,5)\nn
&&~~~~~~~~~~~~~~~~-(\epsilon_1\cdot k_2)\,(\epsilon_3\cdot k_{12})\,(\epsilon_5\cdot k_1)\,\sum_{\shuffle}\,{\cal A}^{\rm BAS}_5(1,\{2,3\}\shuffle5,4|1,2,3,4,5)\Big]\,,
\eea
and
\bea
&&{\rm tr}({\rm F}_{234})\,{\cal A}^{\rm YMS}_5(2,3,4;\underline{1,5}|1,2,3,4,5)\,\xrightarrow[]{\eref{condi-YM-5pt}}\nn
&&~~~~~~~~~~~~~~~~{\rm tr}({\rm F}_{234})\,\Big[(\epsilon_1\cdot k_2)\,(\epsilon_5\cdot k_1)\,\sum_{\shuffle'}\,\sum_{\shuffle}\,{\cal A}^{\rm BAS}_5(1,2\shuffle'3\shuffle5,4|1,2,3,4,5)\nn
&&~~~~~~~~~~~~~~~~+(\epsilon_1\cdot k_{23})\,(\epsilon_5\cdot k_1)\,\sum_{\shuffle}\,{\cal A}^{\rm BAS}_5(1,\{3,2\}\shuffle5,4|1,2,3,4,5)\Big]\,.
\eea

Since all terms in \eref{5ptYM-step1} can be turned to formulas satisfying the reduced expansion \eref{expan-YM+2-reduced}, we can then use the argument after \eref{expan-YM+2-reduced} to conclude the hidden zero for the $5$-point amplitude ${\cal A}^{F^3}_5(1,2,3,4,5)$.

\subsection{General proof}
\label{subsec-proof-YM+2}

In this subsection, we give a general proof for the hidden zero \eref{zero-YM+2}, based on the universal expansion and KK relation.

To begin with, we use \eref{YM+2-toYMS} to expand ${\cal A}^{F^3}_n(\hat{i},\pmb A,\hat{j},\pmb B)$ into YM$\oplus$BAS
amplitudes,
\bea
{\cal A}_n^{F^3}(\hat{i},\pmb A,\hat{j},\pmb B)=&&\sum_{{\pmb{\rho}},\hat{i}\not\in\pmb\rho}\,
{\rm tr}({\rm F}_{{\pmb{\rho}}})\,{\cal A}_n^{\rm YMS}({\pmb{\rho}};{\rm G}_n\setminus\rho|\hat{i},\pmb A,\hat{j},\pmb B)\nn
\,\xrightarrow[]{\eref{condi0-YM}}&&\sum_{{\pmb{\rho}^A},\hat{i}\not\in\pmb\rho^A}\,
{\rm tr}({\rm F}_{{\pmb{\rho}}^A})\,{\cal A}_n^{\rm YMS}({\pmb{\rho}}^A;{\rm G}_n\setminus\rho^A|\hat{i},\pmb A,\hat{j},\pmb B)\nn
+&&\sum_{{\pmb{\rho}^{jA}},\hat{i}\not\in\pmb\rho^{jA}}\,
{\rm tr}({\rm F}_{{\pmb{\rho}}^{jA}})\,{\cal A}_n^{\rm YMS}({\pmb{\rho}}^{jA};{\rm G}_n\setminus\rho^{jA}|\hat{i},\pmb A,\hat{j},\pmb B)\nn
+&&\sum_{{\pmb{\rho}^B},\hat{i}\not\in\pmb\rho^B}\,
{\rm tr}({\rm F}_{{\pmb{\rho}}^B})\,{\cal A}_n^{\rm YMS}({\pmb{\rho}}^B;{\rm G}_n\setminus\rho^B|\hat{i},\pmb A,\hat{j},\pmb B)\nn
+&&\sum_{{\pmb{\rho}^{jB}},\hat{i}\not\in\pmb\rho^{jB}}\,
{\rm tr}({\rm F}_{{\pmb{\rho}}^{jB}})\,{\cal A}_n^{\rm YMS}({\pmb{\rho}}^{jB};{\rm G}_n\setminus\rho^{jB}|\hat{i},\pmb A,\hat{j},\pmb B)\,,~~\label{YM+2-toYMS-for0}
\eea
where the fiducial gluon is chosen to be $g=\hat{i}$. The kinematic condition \eref{condi0-YM} together with the choice of fiducial gluon
imply that the unordered sets $\rho$ can be divided into four sectors: (1) $\rho^A=A^{\rm sub}_1$; (2) $\rho^B=B^{\rm sub}_1$; (3) $\rho^{jA}=\hat{j}\cup A^{\rm sub}_1$; (4) $\rho^{jB}=\hat{j}\cup B^{\rm sub}_1$. Here $A^{\rm sub}_1$ denotes a subset of $A$, while $B^{\rm sub}_1$ denotes a subset of $B$. Suppose elements from $A$ and $B$ enter $\rho$ simultaneously, the factor ${\rm tr}({\rm F}_{\pmb\rho})$ vanishes due to the kinematic condition \eref{condi0-YM}.

The next step is to expand YM$\oplus$BAS amplitudes in \eref{YM+2-toYMS-for0} into BAS KK basis, with $(i,j)=(\hat{i},\hat{j})$ fixed at two ends in the left orderings. We focus on cases $\rho^A=A^{\rm sub}_1$ and $\rho^{jA}=\hat{j}\cup A^{\rm sub}_1$. The treatment for the remaining two cases is analogous.

For $\rho^{jA}=\hat{j}\cup A^{\rm sub}_1$, we first use the cyclic equivalence to arrange the elements in each $\pmb\rho^{jA}$ as $\pmb\rho^{jA}=\{q_1,\pmb\rho_{|\rho|-2},\hat{j}\}$, where $\pmb\rho_{|\rho|-2}$ is the corresponding ordered set for elements in $\rho^{jA}\setminus\{q_1,\hat{j}\}$. Then, we choose the fiducial leg as $f=\hat{i}$, and use \eref{expan-YMS} to expand ${\cal A}_n^{\rm YMS}({\pmb{\rho}}^{jA};{\rm G}_n\setminus\rho^{jA}|\hat{i},\pmb A,\hat{j},\pmb B)$ as
\bea
&&{\cal A}_n^{\rm YMS}({\pmb{\rho}}^{jA};{\rm G}_n\setminus\rho^{jA}|\hat{i},\pmb A,\hat{j},\pmb B)\nn
&=&\sum_{\pmb\gamma^i}\,\sum_{\shuffle'}\,
{\rm E}_{\pmb\gamma^i}\,{\cal A}^{\rm YMS}_n(q_1,\pmb\rho_{|\rho|-2}\shuffle'\pmb\gamma^i,\hat{j};{\rm G}_n\setminus\{\rho^{jA},\gamma^i\}|\hat{i},\pmb A,\hat{j},\pmb B)\nn
&\xrightarrow[]{\eref{condi0-YM}}&\sum_{\pmb\gamma^{iA}}\,\sum_{\shuffle'}\,
{\rm E}_{\pmb\gamma^{iA}}\,{\cal A}^{\rm YMS}_n(q_1,\pmb\rho_{|\rho|-2}\shuffle'\pmb\gamma^{iA},\hat{j};{\rm G}_n\setminus\{\rho^{jA},\gamma^{iA}\}|\hat{i},\pmb A,\hat{j},\pmb B)\nn
&=&\sum_{\pmb\eta^A_1}\,\sum_{\pmb\eta^A_2}\,{\rm E}_{\pmb\gamma^{iA}}\,{\cal A}^{\rm YMS}_n(\pmb\eta^A_1,\hat{i},\pmb\eta^A_2,\hat{j};{\rm G}_n\setminus\{\rho^{jA},\gamma^{iA}\}|\hat{i},\pmb A,\hat{j},\pmb B)\nn
&=&\sum_{\pmb\eta^A_1}\,\sum_{\pmb\eta^A_2}\,(-)^{|\eta_1^A|}\,{\rm E}_{\pmb\gamma^{iA}}\,\sum_{\shuffle''}\,{\cal A}^{\rm YMS}_n(\hat{i},\pmb\eta^A_2\shuffle''\pmb\eta_1^{AT},\hat{j};{\rm G}_n\setminus\{\rho^{jA},\gamma^{iA}\}|\hat{i},\pmb A,\hat{j},\pmb B)\,.~~\label{rho-jA}
\eea
In the above, the second step uses the observation that the kinematic condition \eref{condi0-YM} requires the effective $\pmb\gamma^i$ to satisfy $\gamma^{iA}=\hat{i}\cup A^{\rm sub}_2$, since each $Y_{r_1}$ in ${\rm E}_{\pmb\gamma^i}$ receives contributions solely from $A$. In the third step, we have relabeled each ordering $(q_1,\pmb\rho_{|\rho|-2}\shuffle'\pmb\gamma^{iA},\hat{j})$ as $(\pmb\eta^A_1,\hat{i},\pmb\eta^A_2,\hat{j})$, since $\hat{i}\in\gamma^{iA}$. Elements in ordered sets $\pmb\eta^A_1$ and $\pmb\eta^A_2$ are solely from $A$, as indicated by the superscript $A$. Notice that when summing over $\pmb\eta^A_1$ and $\pmb\eta^A_2$, the ordered sets $\pmb\gamma^{iA}$ and shuffles $\shuffle'$ are implicitly summed. The final step uses the generalized KK relation \eref{KK-generalized} to convert YM$\oplus$BAS amplitudes to those with $(\hat{i},\hat{j})$ fixed at two ends in the left orderings, where $\pmb\eta_1^{AT}$ stands for the inverse of $\pmb\eta_1^A$.

To proceed, one can iteratively apply the expansion \eref{expan-YMS}, to further expand the YM$\oplus$BAS amplitudes in the final step of \eref{rho-jA}. Repeating this manipulation creates a series of ordered sets $\pmb\gamma_1,\,\pmb\gamma_2,\,\pmb\gamma_3,\cdots$. Here we have omitted the superscripts of these $\pmb\gamma_\ell$ since the choices of fiducial legs are irrelevant. The kinematic condition \eref{condi0-YM} implies that the elements in each $\pmb\gamma_\ell$ are solely from $A$ or $B$, otherwise the corresponding ${\rm E}_{\pmb\gamma_\ell}$ will vanish. In other words, elements from $A$ and $B$ cannot enter any individual $\pmb\gamma_\ell$ simultaneously. Thus, we can divide these $\pmb\gamma_\ell$ into two sectors $\{\pmb\gamma_1^A,\pmb\gamma_2^A,\cdots\}$ and $\{\pmb\gamma_1^B,\pmb\gamma_2^B,\cdots\}$, where each $\pmb\gamma_{\ell}^A$ contains elements from $A$, while each $\pmb\gamma_{\ell}^B$ contains elements from $B$. Each YM$\oplus$BAS amplitude in the final step of \eref{rho-jA} is then expanded into BAS KK basis as
\bea
&&{\cal A}^{\rm YMS}_n(\hat{i},\pmb\eta^A_2\shuffle''\pmb\eta_1^{AT},\hat{j};{\rm G}_n\setminus\{\rho^{jA},\gamma^{iA}\}|\hat{i},\pmb A,\hat{j},\pmb B)\nn
&\xrightarrow[]{\eref{condi0-YM}}&\Big[\sum_{\{\pmb\gamma_{\ell}^A\}}\,\prod_{\ell}\,\Big(\sum_{\shuffle_{\ell}}\,{\rm E}_{\pmb\gamma^A_{\ell}}\Big)\Big]\,
\Big[\sum_{\{\pmb\gamma^B_{\ell'}\}}\,\prod_{\ell'}\,\Big(\sum_{\W\shuffle_{\ell'}}\,{\rm E}_{\pmb\gamma^B_{\ell'}}\Big)\Big]\nn
&&{\cal A}^{\rm BAS}_n(\hat{i},\pmb\eta^A_2\shuffle''\pmb\eta_1^{AT}\shuffle_{1}\pmb\gamma^A_{1}\shuffle_{2}\pmb\gamma^A_{2}\cdots\W\shuffle_{1}\pmb\gamma^B_{1}
\W\shuffle_{2}\pmb\gamma^B_{2}\cdots,\hat{j}|\hat{i},\pmb A,\hat{j},\pmb B)\nn
&\xrightarrow[]{\eref{condi0-YM}}&\Big[\sum_{\{\pmb\gamma^A_{\ell}\}}\,\prod_{\ell}\,\Big(\sum_{\shuffle_{\ell}}\,{\rm E}_{\pmb\gamma^A_{\ell}}\Big)\Big]\,
\Big[\sum_{\{\pmb\gamma^B_{\ell'}\}}\,{\rm E}_{\pmb\gamma^B_{1}}\,\prod_{\ell'\neq1}\,\Big(\sum_{\W\shuffle_{\ell'}}\,{\rm E}_{\pmb\gamma^B_{\ell'}}\Big)\Big]\,
\sum_{\shuffle}\nn
&&{\cal A}^{\rm BAS}_n(\hat{i},\{\pmb\eta^A_2\shuffle''\pmb\eta_1^{AT}\shuffle_{1}\pmb\gamma^A_{1}\shuffle_{2}\pmb\gamma^A_{2}\cdots\}\shuffle\{\pmb\gamma^B_{1}
\W\shuffle_{2}\pmb\gamma^B_{2}\cdots\},\hat{j}|\hat{i},\pmb A,\hat{j},\pmb B)\,,~~\label{reducedexpan-rhojA}
\eea
where the associative law for shuffles is used in the last step. In the above, the summation over $\{\pmb\gamma^A_{\ell}\}$ is for divisions which split the remaining elements of $A$ into ordered sets $\pmb\gamma^A_{\ell}$, and the summation over $\{\pmb\gamma^B_{\ell'}\}$ is understood analogously. Notice that $\{\pmb\gamma^A_{\ell}\}$ and $\pmb\gamma^B_{\ell'}$ are two ordered sets, where each $\pmb\gamma^A_{\ell}$ or $\pmb\gamma^B_{\ell'}$ serves as an element. That is, each step of iteratively expansion creates only one $\pmb\gamma^A_\ell$ or $\pmb\gamma^B_{\ell'}$, and the ordering of steps determines the ordering of elements in $\{\pmb\gamma^A_\ell\}$ and $\{\pmb\gamma^B_{\ell'}\}$ \footnote{To achieve the formula \eref{reducedexpan-rhojA}, one direct way is to arrange the ordering of steps as follows: first, create $\pmb\gamma^A_\ell$ one by one, and then create $\pmb\gamma^B_{\ell'}$ one by one. However, since all kinematic factors are independent of the shuffles $\shuffle$ in the final step of \eref{reducedexpan-rhojA}, the order of creating $\pmb\gamma^A_\ell$ and $\pmb\gamma^B_{\ell'}$ is irrelevant, and the only important orderings are those in $\{\pmb\gamma^A_\ell\}$ and $\{\pmb\gamma^B_{
\ell'}\}$, respectively.}. Clearly, the corresponding kinematic factors ${\rm E}_{\pmb\gamma^A_{\ell}}$ or ${\rm E}_{\pmb\gamma^B_{\ell'}}$ will be altered if we modify the ordering of elements in $\{\pmb\gamma^A_{\ell}\}$ or $\{\pmb\gamma^B_{\ell'}\}$. The products of $\sum_{\shuffle_\ell}{\rm E}_{\pmb\gamma^A_\ell}$ and $\sum_{\W\shuffle_\ell'}{\rm E}_{\pmb\gamma^B_{\ell'}}$ are understood as
\bea
&&\prod_{\ell}\,\Big(\sum_{\shuffle_{\ell}}\,{\rm E}_{\pmb\gamma^A_{\ell}}\Big)~~{\rm means}~~\sum_{\shuffle_1}\,\sum_{\shuffle_2}\,\sum_{\shuffle_3}\,\cdots\,{\rm E}_{\pmb\gamma^A_{1}}\,{\rm E}_{\pmb\gamma^A_{2}}\,{\rm E}_{\pmb\gamma^A_{2}}\,\cdots\,,\nn
&&\prod_{\ell'}\,\Big(\sum_{\W\shuffle_{\ell'}}\,{\rm E}_{\pmb\gamma^B_{\ell'}}\Big)~~{\rm means}~~\sum_{\W\shuffle_1}\,\sum_{\W\shuffle_2}\,\sum_{\W\shuffle_3}\,\cdots\,{\rm E}_{\pmb\gamma^B_{1}}\,{\rm E}_{\pmb\gamma^B_{2}}\,{\rm E}_{\pmb\gamma^B_{2}}\,\cdots\,.~~\label{understood-prod}
\eea

Because of the kinematic condition \eref{condi0-YM} and the definition of ${\rm E}_{\pmb\gamma^f}$ in \eref{define-El}, the effective part of $Y_{r_1}$ in any ${\rm E}_{\pmb\gamma^B_{\ell'}}$ in \eref{reducedexpan-rhojA} does not receive contributions from $A$, while the effective part of $Y_{r_1}$ in any ${\rm E}_{\pmb\gamma^A_{\ell}}$ does not receive contributions from $B$. It means, these ${\rm E}_{\pmb\gamma^A_{\ell}}$ and ${\rm E}_{\pmb\gamma^B_{\ell'}}$ are independent of the shuffles denoted as $\shuffle$ in the final step of \eref{reducedexpan-rhojA}. This is the reason why we put $\sum_\shuffle$ after ${\rm E}_{\pmb\gamma^A_{\ell}}$ and ${\rm E}_{\pmb\gamma^B_{\ell}}$. Meanwhile, factors ${\rm tr}(F_{\pmb\rho^{jA}})$ in \eref{YM+2-toYMS-for0} and ${\rm E}_{\pmb\gamma^{iA}}$ in \eref{rho-jA} are clearly independent of these $\shuffle$. Consequently, the $\pmb\rho^{jA}$-part in \eref{YM+2-toYMS-for0} satisfies the reduced expansion \eref{expan-YM+2-reduced}, that is, this part can be expanded into the BAS KK basis with $(\hat{i},\hat{j})$ fixed at two ends in the left orderings, such that the coefficients are independent of the shuffles $\shuffle$ between $\pmb A'$ and $\pmb B'$. In \eref{reducedexpan-rhojA}, such $\pmb A'$ are given by $\pmb\eta^A_2\shuffle''\pmb\eta_1^{AT}\shuffle_{1}\pmb\gamma^A_{1}\shuffle_{2}\pmb\gamma^A_{2}\cdots$, while $\pmb B'$
are given by $\pmb\gamma^B_{1}
\W\shuffle_{2}\pmb\gamma^B_{2}\cdots$.

Now we turn to the $\pmb\rho^A$-part in \eref{YM+2-toYMS-for0}. For any $\pmb\rho^A=\{q_1,\pmb\rho_{|\rho|-2},q_{|\rho|}\}$, we first choose the fiducial leg as $f=\hat{j}$, and expand this part as
\bea
&&{\cal A}_n^{\rm YMS}({\pmb{\rho}}^A;{\rm G}_n\setminus\rho^A|\hat{i},\pmb A,\hat{j},\pmb B)\nn
&=&\sum_{\pmb\gamma^j}\,\sum_{\shuffle'}\,{\rm E}_{\pmb\gamma^j}\,{\cal A}^{\rm YMS}_n(q_1,\pmb\rho_{|\rho|-2}\shuffle'\pmb\gamma^j,q_{|\rho|};{\rm G}_n\setminus\{\rho^A,\gamma^j\}|\hat{i},\pmb A,\hat{j},\pmb B)\nn
&=&\sum_{\pmb\gamma^j_{\not{i}}}\,\sum_{\shuffle'}\,{\rm E}_{\pmb\gamma^j_{\not{i}}}\,{\cal A}^{\rm YMS}_n(q_1,\pmb\rho_{|\rho|-2}\shuffle'\pmb\gamma^j_{\not{i}},q_{|\rho|};{\rm G}_n\setminus\{\rho^A,\gamma^j_{\not{i}}\}|\hat{i},\pmb A,\hat{j},\pmb B)\nn
&&+\sum_{\pmb\gamma^j_i}\,\sum_{\shuffle'}\,{\rm E}_{\pmb\gamma^j_i}\,{\cal A}^{\rm YMS}_n(q_1,\pmb\rho_{|\rho|-2}\shuffle'\pmb\gamma^j_i,q_{|\rho|};{\rm G}_n\setminus\{\rho^A,\gamma^j_i\}|\hat{i},\pmb A,\hat{j},\pmb B)\,,
~~\label{2cases}
\eea
where we have separated $\gamma^j$ into two sectors which are: (1) $\gamma^j_{\not{i}}$ satisfying $i\not\in\gamma^j_{\not{i}}$; (2) $\gamma^j_i$ satisfying $i\in\gamma^j_i$. For either $\gamma^j_{\not{i}}$ or $\gamma^j_i$, the combinatorial momentum $Y_{\gamma_1}$ contains external momenta solely from $A$. This observation together with the kinematic condition \eref{condi0-YM} yield strong constraints on elements in $\gamma^j_{\not{i}}$ or $\gamma^j_i$. For the first part, they imply that the elements in $\gamma^j_{\not{i}}$ are solely from $A$. Therefore, the corresponding YMS$\oplus$BAS amplitudes in \eref{2cases} share the structure of ${\cal A}_n^{\rm YMS}({\pmb{\rho}}^{jA};{\rm G}_n\setminus\rho^{jA}|\hat{i},\pmb A,\hat{j},\pmb B)$ in the first line of \eref{rho-jA}, namely $\hat{j}$ is a scalar while $\hat{i}$ is a gluon, and all the remaining scalars are from $A$. Thus one can repeat the previous process to achieve the formula \eref{reducedexpan-rhojA}.
For the second part---the $\gamma^j_i$-part, they imply two possibilities.
The first one, the elements in $\gamma^j_i\setminus\{\hat{i},\hat{j}\}$ are solely from $A$. For this case, the corresponding YM$\oplus$BAS amplitudes in \eref{2cases} share the structure of ${\cal A}^{\rm YMS}_n(q_1,\pmb\rho_{|\rho|-2}\shuffle_1\pmb\gamma^{iA},\hat{j};{\rm G}_n\setminus\{\rho^{jA},\gamma^{iA}\}|\hat{i},\pmb A,\hat{j},\pmb B)$ in the third line of \eref{rho-jA}, that is, both of $\hat{i}$ and $\hat{j}$ are scalars, and the remaining scalars are from $A$. Therefore, the same process can be repeated again, yielding the formula \eref{reducedexpan-rhojA}. Therefore, we only need to consider the second case $\pmb\gamma^j_i=\{\pmb\gamma^A,\hat{i},\pmb\gamma^B,\hat{j}\}$,
where $\pmb\gamma^A$ and $\pmb\gamma^B$ contain elements from $A$ and $B$, respectively.

For $\pmb\gamma^j_i=\{\pmb\gamma^A,\hat{i},\pmb\gamma^B,\hat{j}\}$, let us reorganize the YM$\oplus$BAS amplitudes in the last line of \eref{2cases}
as
\bea
&&\sum_{\shuffle'}\,{\rm E}_{\pmb\gamma^j_i}\,{\cal A}^{\rm YMS}_n(q_1,\pmb\rho_{|\rho|-2}\shuffle'\pmb\gamma^j_i,q_{|\rho|};{\rm G}_n\setminus\{\rho^A,\gamma^j_i\}|\hat{i},\pmb A,\hat{j},\pmb B)\nn
&=&\sum_{\pmb\eta^A_1}\,\sum_{\pmb\eta^A_2}\,{\rm E}_{\pmb\gamma^j_i}\,\sum_{\shuffle}\,{\cal A}^{\rm YMS}_n(\pmb\eta^A_1,\hat{i},\pmb\eta^A_2\shuffle\pmb\gamma^B,\hat{j};{\rm G}_n\setminus\{\rho^A,\gamma^j_i\}|\hat{i},\pmb A,\hat{j},\pmb B)\nn
&=&\sum_{\pmb\eta^A_1}\,\sum_{\pmb\eta^A_2}\,(-)^{|\eta^A_1|}\,{\rm E}_{\pmb\gamma^j_i}\,\sum_{\shuffle}\,\sum_{\shuffle''}\,{\cal A}^{\rm YMS}_n(\hat{i},\pmb\eta^{AT}_1\shuffle''\pmb\eta^A_2\shuffle\pmb\gamma^B,\hat{j};{\rm G}_n\setminus\{\rho^A,\gamma^j_i\}|\hat{i},\pmb A,\hat{j},\pmb B)\,,~~\label{case-AB}
\eea
where $\eta^A_1\cup\eta^A_2=\rho^A\cup\gamma^A$. The first equality is based on the observation that each ordering $(q_1,\pmb\rho_{|\rho|-2}\shuffle_1\pmb\gamma^j_i,q_{|\rho|})$ in the first line of \eref{case-AB} can be characterized as
$(q_1,\cdots,\hat{i},\pmb\eta^A_2\shuffle\pmb\gamma^B,\hat{j},\cdots,q_{|\rho|})$, where $\pmb\eta^A_2$ is a part of $\pmb\rho_{|\rho|-2}$, namely $\pmb\rho_{|\rho|-2}=\{\cdots,\pmb\eta^A_2,\cdots\}$. Thus one can use the cyclic equivalence to move $\hat{j}$ to the right end, and replacing the summation over $\shuffle'$ by summations over $\pmb\eta^A_1$, $\pmb\eta^A_2$ and $\shuffle$ arising from $\shuffle'$. Notice that each coefficient ${\rm E}_{\pmb\gamma^j_i}$ is independent of $\shuffle$, as implied by the definition.
The final equality uses the generalized KK relation \eref{KK-generalized} to convert orderings $(\pmb\eta^A_1,\hat{i},\pmb\eta^A_2\shuffle\pmb\gamma^B,\hat{j})$ to new ones $(\hat{i},\pmb\eta^{AT}_1\shuffle''\pmb\eta^A_2\shuffle\pmb\gamma^B,\hat{j})$. Similar as in \eref{reducedexpan-rhojA}, one can further expand YM$\oplus$BAS amplitudes, obtaining
\bea
&&\sum_{\shuffle}\,{\cal A}^{\rm YMS}_n(\hat{i},\pmb\eta^{AT}_1\shuffle''\pmb\eta^A_2\shuffle\pmb\gamma^B,\hat{j};{\rm G}_n\setminus\{\rho^A,\gamma^j_i\}|\hat{i},\pmb A,\hat{j},\pmb B)\nn
&\xrightarrow[]{\eref{condi0-YM}}&\sum_{\shuffle}\,\Big[\sum_{\{\pmb\gamma^A_{\ell}\}}\,\prod_{\ell}\,\Big(\sum_{\shuffle_{\ell}}\,{\rm E}_{\pmb\gamma^A_{\ell}}\Big)\Big]\,
\Big[\sum_{\{\pmb\gamma^B_{\ell'}\}}\,\prod_{\ell'}\,\Big(\sum_{\W\shuffle_{\ell'}}\,{\rm E}_{\pmb\gamma^B_{\ell'}}\Big)\Big]\nn
&&{\cal A}^{\rm BAS}_n(\hat{i},\pmb\eta^{AT}_1\shuffle''\pmb\eta^A_2\shuffle\pmb\gamma^B\shuffle_{1}\pmb\gamma^A_{1}\shuffle_{2}\pmb\gamma^A_{2}\cdots
\W\shuffle_{1}\pmb\gamma^B_{1}
\W\shuffle_{2}\pmb\gamma^B_{2}\cdots,\hat{j}|\hat{i},\pmb A,\hat{j},\pmb B)\nn
&\xrightarrow[]{\eref{condi0-YM}}&\Big[\sum_{\{\pmb\gamma^A_{\ell}\}}\,\prod_{\ell}\,\Big(\sum_{\shuffle_{\ell}}\,{\rm E}_{\pmb\gamma^A_{\ell}}\Big)\Big]\,
\Big[\sum_{\{\pmb\gamma^B_{\ell'}\}}\,\prod_{\ell'}\,\Big(\sum_{\W\shuffle_{\ell'}}\,{\rm E}_{\pmb\gamma^B_{\ell'}}\Big)\Big]\,
\sum_{\shuffle}\nn
&&{\cal A}^{\rm BAS}_n(\hat{i},\{\pmb\eta^{AT}_1\shuffle''\pmb\eta^A_2\shuffle_{1}\pmb\gamma^A_{1}\shuffle_{2}\pmb\gamma^A_{2}\cdots\}\shuffle\{\pmb\gamma^B
\W\shuffle_{1}\pmb\gamma^B_{1}
\W\shuffle_{2}\pmb\gamma^B_{2}\cdots\},\hat{j}|\hat{i},\pmb A,\hat{j},\pmb B)\,.~~\label{reducedexpan-rhoA}
\eea
Each ${\rm E}_{\pmb\gamma^A_{\ell}}$ or ${\rm E}_{\pmb\gamma^B_{\ell'}}$ in \eref{reducedexpan-rhoA}, as well as ${\rm tr}({\rm F}_{\pmb\rho^A})$ and ${\rm E}_{\pmb\gamma^j_i}$ in \eref{YM+2-toYMS-for0} and \eref{case-AB}, are clearly independent of the shuffles labeled by $\shuffle$ in \eref{reducedexpan-rhoA}. Thus, the final part with $\pmb\gamma^j_i=\{\pmb\gamma^A,\hat{i},\pmb\gamma^B,\hat{j}\}$ also satisfies the reduced expansion \eref{expan-YM+2-reduced}.

The manipulations for the $\pmb\rho^{jB}$-part and $\pmb\rho^B$-part in \eref{YM+2-toYMS-for0} are performed in parallel. Consequently, all four parts in \eref{YM+2-toYMS-for0} satisfy the reduced expansion \eref{expan-YM+2-reduced}. It follows that one can apply the KK relation \eref{KK} to transform all BAS amplitudes in the reduced expansion to those with compatible orderings defined in \eref{compa-order}. Ultimately, the hidden zero for each $F^3$ amplitude can be interpreted in terms of the hidden zeros for BAS amplitudes with compatible orderings.

\section{Hidden zeros for higher-derivative GR amplitudes}
\label{sec-zero-GR}

In this section, we extend the investigation of hidden zeros to $R^2$ and $R^3$ amplitudes using a similar approach based on universal expansions. As noted in section \ref{sec-intro}, the kinematic condition for hidden zeros leads to unavoidable singular propagators in unordered graviton amplitudes.
A preliminary way to address this issue is to recognize that each propagator inherently carries an infinitesimal imaginary part $i\epsilon$ in its denominator. Using this prescription, we establish the existence of hidden zeros for $R^2$ and $R^3$ amplitudes in subsection \ref{subsec-zero-GRG3}.

However, the reliance on the $i\epsilon$ argument is not fully satisfactory. To place the conclusion on a more rigorous foundation, we reveal a systematic cancellation mechanism that inherently removes all divergences, yielding a finite effective expansion. The existence of such cancellations allows us to establish the hidden zeros for $R^2$ and $R^3$ amplitudes unambiguously. In subsection \ref{subsec-cancel-mecha}, we describe the main mechanism and conclusion of the cancellations, especially the behavior of effective coefficients in \eref{cancel-toprove} and its major consequence in \eref{deduce}. An explicit example and a general proof are then provided in subsequent subsections \ref{subsec-cancel-example} and \ref{subsec-cancel-proof}, respectively.

\subsection{Hidden zeros for $R^2$ and $R^3$ amplitudes}
\label{subsec-zero-GRG3}

The study in the previous section shows that the expansion of $F^3$ amplitudes in \eref{YM+2toKK} can be reduced to \eref{expan-YM+2-reduced}, on the special loci \eref{condi0-YM} in kinematic space. As a direct generalization, the expansion of $R^3$ amplitudes in \eref{expan-R3} can also be reduced to
\bea
{\cal A}_n^{R^3}&\xrightarrow[]{\eref{condi0-YM}}&\sum_{\pmb A}\,\sum_{\pmb B}\,\sum_{\pmb A'}\,\sum_{\pmb B'}\,C_{F^3}(\epsilon,\pmb A,\pmb B)\,C_{F^3}(\W\epsilon,\pmb A',\pmb B')\,\sum_{\shuffle}\,\sum_{\shuffle'}\,{\cal A}^{\rm BAS}_n(\hat{i},\pmb A\shuffle\pmb B,\hat{j}|\hat{i},\pmb A'\shuffle'\pmb B',\hat{j})\,,~~\label{expan-R3-reduced}
\eea
due to the double copy construction in \eref{CHY-integrand} and the expansion of partial integrands in \eref{expan-P}. Then, by applying the KK relation \eref{KK} to left and right orderings of BAS amplitudes in \eref{expan-R3-reduced}, we straightforwardly obtain the hidden zeros for $n$-point $R^3$ amplitudes,
\bea
{\cal A}^{R^3}_n&\xrightarrow[]{\eref{condi0-YM}}&0\,,~~\label{zero-R3}
\eea
follow from the hidden zeros for BAS amplitudes in \eref{zero-BAS}. Notice that the kinematic condition \eref{condi0-YM} also holds for another copy of polarization vectors $\{\W\epsilon_\ell\}$, since $\epsilon_\ell^\mu=\W\epsilon_\ell^\mu$.

However, the above naive argument has a serious obstacle caused by divergent propagators. That is, each propagator $1/s_{ab}$ with $a\in A$
and $b\in B$ is divergent under the condition \eref{condi0-YM}. In the ordered $F^3$ case, such divergent propagators are forbidden by the orderings $(\hat{i},\pmb A,\hat{j},B)$ which are compatible with the hidden zero condition. In contrast, for the current unordered $R^3$ case, these divergent propagators are unavoidable. As can be seen in the previous section, the reduced expansion \eref{expan-YM+2-reduced} is achieved by removing kinematic factors vanishing under the kinematic condition \eref{condi0-YM}. However, since BAS amplitudes in the general expansion formula in \eref{expan-R3} may contain divergences, one cannot naively drop the vanishing kinematic factors to obtain the reduced expansion \eref{expan-R3-reduced}.

To solve this problem, a simple argument is to notice that each propagator always contains an infinitesimal imaginary part $i\epsilon$ in its denominator. Therefore, the divergent propagators are $1/i\epsilon$, while the vanishing kinematic factors in numerators are exactly zero. This observation allows us to repeat the reduction in the previous section, obtain the expansion in \eref{expan-R3-reduced}, and thereby establish the existence of the hidden zeros. The above argument is not the most satisfactory one. In subsequent subsections, we will demonstrate systematic cancellations of divergences without referring to the imaginary part $i\epsilon$, thereby ensuring the validity of the reduced expansion \eref{expan-R3-reduced}.

The $R^2$ amplitudes have the similar hidden zeros,
\bea
{\cal A}^{R^2}_n&\xrightarrow[]{\eref{condi0-YM}}&0\,,~~\label{zero-R2}
\eea
based on the observation that the expansion of usual YM amplitudes can also be reduced to
\bea
{\cal A}_n^{\rm YM}(\pmb\sigma_n)&\xrightarrow[]{\eref{condi0-YM}}&\sum_{\pmb A'}\,\sum_{\pmb B'}\,C_{\rm YM}(\epsilon,\pmb A',\pmb B')\,\sum_{\shuffle}\,{\cal A}^{\rm BAS}_n(\hat{i},\pmb A'\shuffle\pmb B',\hat{j}|\pmb\sigma_n)\,.~~\label{expan-YM-reduced}
\eea
To get the above reduced expansion, we choose $(i,j)=(\hat{i},\hat{j})$ in \eref{YMtoYMS}, and observe that the kinematic condition \eref{condi0-YM}
forces that the elements in any $\pmb\pi$ should be solely from $A$ or $B$. Thus we obtain
\bea
{\cal A}^{\rm YM}_n(\pmb\sigma_n)&\xrightarrow[]{\eref{condi0-YM}}&\sum_{\pmb\pi^A,0\leq|\pi^A|\leq |A|}\,{\rm T}^{(\hat{i},\hat{j})}_{\pmb\pi^A}\,{\cal A}^{\rm YMS}_n(\hat{i},\pmb\pi^A,\hat{j};{\rm G}_n\setminus\{\hat{i},\hat{j},\pi^A\}|\pmb\sigma_n)\nn
&&+\sum_{\pmb\pi^B,0\leq|\pi^B|\leq |B|}\,{\rm T}^{(\hat{i},\hat{j})}_{\pmb\pi^B}\,{\cal A}^{\rm YMS}_n(\hat{i},\pmb\pi^B,\hat{j};{\rm G}_n\setminus\{\hat{i},\hat{j},\pi^B\}|\pmb\sigma_n)\,,~~\label{YMtoYMS-separated}
\eea
where $\pi^A\subset A$ and $\pi^B\subset B$. Then, we recursively use the expansion \eref{expan-YMS} to achieve
\bea
{\cal A}^{\rm YM}_n(\pmb\sigma_n)&\xrightarrow[]{\eref{condi0-YM}}&
\sum_{\pmb\pi^A,0\leq|\pi^A|\leq |A|}\,{\rm T}^{(\hat{i},\hat{j})}_{\pmb\pi^A}\,\Big[\sum_{\{\pmb\gamma^A_{\ell}\}}\,\prod_{\ell}\,\Big(\sum_{\shuffle_{\ell}}\,{\rm E}_{\pmb\gamma^A_{\ell}}\Big)\Big]\,\Big[\sum_{\{\pmb\gamma^B_{\ell'}\}}\,\prod_{\ell'}\,\Big(\sum_{\W\shuffle_{\ell'}}\,{\rm E}_{\pmb\gamma^B_{\ell'}}\Big)\Big]\nn
&&~~~~~~~~~~~~{\cal A}^{\rm YMS}_n(\hat{i},\pmb\pi^A\shuffle_{1}\pmb\gamma^A_{1}\shuffle_{2}\pmb\gamma^A_{2}\cdots
\W\shuffle_{1}\pmb\gamma^B_{1}\W\shuffle_{2}\pmb\gamma^B_{2}\cdots,
\hat{j}|\pmb\sigma_n)\nn
&&+\sum_{\pmb\pi^B,0\leq|\pi^B|\leq |B|}\,{\rm T}^{(\hat{i},\hat{j})}_{\pmb\pi^B}\,\Big[\sum_{\{\W{\pmb\gamma}^A_{\ell}\}}\,\prod_{\ell}\,\Big(\sum_{\shuffle_{\ell}}\,{\rm E}_{\W{\pmb\gamma}^A_{\ell}}\Big)\Big]\,\Big[\sum_{\{\W{\pmb\gamma}^B_{\ell'}\}}\,\prod_{\ell'}\,\Big(\sum_{\W\shuffle_{\ell'}}\,{\rm E}_{\W{\pmb\gamma}^B_{\ell'}}\Big)\Big]\nn
&&~~~~~~~~~~~~{\cal A}^{\rm YMS}_n(\hat{i},\pmb\pi^B\W\shuffle_{1}\W{\pmb\gamma}^B_{1}\W\shuffle_{2}\W{\pmb\gamma}^B_{2}\cdots
\shuffle_{1}\W{\pmb\gamma}^A_{1}\shuffle_{2}\W{\pmb\gamma}^A_{2}\cdots
,\hat{j}|\pmb\sigma_n)\nn
&\xrightarrow[]{\eref{condi0-YM}}&
\sum_{\pmb\pi^A,0\leq|\pi^A|\leq |A|}\,{\rm T}^{(\hat{i},\hat{j})}_{\pmb\pi^A}\,\Big[\sum_{\{\pmb\gamma^A_{\ell}\}}\,\prod_{\ell}\,\Big(\sum_{\shuffle_{\ell}}\,{\rm E}_{\pmb\gamma^A_{\ell}}\Big)\Big]\,\Big[\sum_{\{\pmb\gamma^A_{\ell'}\}}\,{\rm E}_{\pmb\gamma^B_{1}}\,\prod_{\ell'\neq1}\,\Big(\sum_{\W\shuffle_{\ell'}}\,{\rm E}_{\pmb\gamma^B_{\ell'}}\Big)\Big]\,\sum_{\shuffle}\nn
&&~~~~~~~~~~~~{\cal A}^{\rm YMS}_n(\hat{i},\{\pmb\pi^A\shuffle_{1}\pmb\gamma^A_{1}\shuffle_{2}\pmb\gamma^A_{2}\cdots\}
\shuffle\{\pmb\gamma^B_{1}\W\shuffle_{2}\pmb\gamma^B_{2}
\cdots\},\hat{j}|\pmb\sigma_n)\nn
&&+\sum_{\pmb\pi^B,0\leq|\pi^B|\leq |B|}\,{\rm T}^{(\hat{i},\hat{j})}_{\pmb\pi^B}\,\Big[\sum_{\{\W{\pmb\gamma}^A_{\ell}\}}\,{\rm E}_{\W{\pmb\gamma}^A_{1}}\,\prod_{\ell\neq1}\,\Big(\sum_{\shuffle_{\ell}}\,{\rm E}_{\W{\pmb\gamma}^A_{\ell}}\Big)\Big]\,\Big[\sum_{\{\W{\pmb\gamma}^B_{\ell'}\}}\,\prod_{\ell'}\,\Big(\sum_{\W\shuffle_{\ell'}}\,{\rm E}_{\W{\pmb\gamma}^B_{\ell'}}\Big)\Big]\,\sum_{\shuffle}\nn
&&~~~~~~~~~~~~{\cal A}^{\rm YMS}_n(\hat{i},\{\pmb\pi^B\W\shuffle_{1}\W{\pmb\gamma}^B_{1}\W\shuffle_{2}\W{\pmb\gamma}^B_{2}\cdots\}
\shuffle\{\W{\pmb\gamma}^A_{1}\shuffle_{2}\W{\pmb\gamma}^A_{2}\cdots
\},\hat{j}|\pmb\sigma_n)\,,~~\label{YMtoKK-result}
\eea
satisfying the reduced expansion \eref{expan-YM-reduced}. The above process is analogous to that in \eref{reducedexpan-rhojA}, thus we omit explanations for each step. The reduced expansions \eref{expan-YM+2-reduced} and \eref{expan-YM-reduced} implies that the expansion of $R^2$ amplitudes in \eref{expan-R2} can be turned to
\bea
{\cal A}_n^{R^2}&\xrightarrow[]{\eref{condi0-YM}}&\sum_{\pmb A}\,\sum_{\pmb B}\,\sum_{\pmb A'}\,\sum_{\pmb B'}\,C_{F^3}(\epsilon,\pmb A,\pmb B)\,C_{\rm YM}(\W\epsilon,\pmb A',\pmb B')\,\sum_{\shuffle}\,\sum_{\shuffle'}\,{\cal A}^{\rm BAS}_n(\hat{i},\pmb A\shuffle\pmb B,\hat{j}|\hat{i},\pmb A'\shuffle'\pmb B',\hat{j})\,,~~\label{expan-R2-reduced}
\eea
thus one can again apply the KK relation \eref{KK} to find the hidden zeros in \eref{zero-R2}.

Again, the above argument for hidden zeros for $R^2$ amplitudes holds as long as the divergent propagators can be bypassed. This difficulty can be solved either by employing the argument based on infinitesimal imaginary parts $i\epsilon$, or by demonstrating the cancellations of divergences which will be studied in the rest of this section.

\subsection{Cancellations of divergences and effective expansions}
\label{subsec-cancel-mecha}

In this subsection, we show the main mechanism and conclusion of the cancellations of divergences, and the reason why such cancellations ensure the validity of the reduced expansions of $R^2$ and $R^3$ amplitudes in \eref{expan-R2-reduced} and \eref{expan-R3-reduced}.

To describe degrees of divergences, let us parameterize each Lorentz invariant in the kinematic condition \eref{condi0-YM} as,
\bea
k_a\cdot k_b\to\tau\,k_a\cdot k_b\,,~~~~k_a\cdot\epsilon_b\to\tau\,k_a\cdot\epsilon_b\,,~~~~
\epsilon_a\cdot k_b\to\tau\,\epsilon_a\cdot k_b\,,~~~~\epsilon_a\cdot\epsilon_b\to\tau\,\epsilon_a\cdot\epsilon_b\,,~~~~~~~~\tau\to0\,,~~\label{condi-tau}
\eea
A power counting in the parameter $\tau$ then characterizes the divergences. Consider an individual Feynman diagram of BAS theory,
which contains divergent propagators $1/s_{a_1b_1}$, $1/s_{a_2b_2}$, ..., $1/s_{a_tb_t}$. These propagators contribute $\tau^{-t}$.
As in \eref{swapp-pq}, \eref{example1-gamma} and \eref{example2-gamma}, we use $\Gamma$ to label the product of massless propagators from this diagram, without the $\pm$ sign.
For each partial BAS amplitude which contains the particular $\Gamma$ at the $\tau^{-t}$ order, each pair of legs $(a_\ell,b_\ell)$ with $\ell\in\{1,\cdots,t\}$ should be adjacent to each other in either of two orderings $(\hat{i},\pmb\a_{n-2},\hat{j})$ and $(\hat{i},\pmb\a'_{n-2},\hat{j})$ in \eref{expan-R3} or \eref{expan-R2}, due to the definition of partial BAS amplitudes. For instance, if $t=1$, $(\hat{i},\pmb\a_{n-2},\hat{j})$ and $(\hat{i},\pmb\a'_{n-2},\hat{j})$ should take the form $(\hat{i},\cdots,a_1,b_1,\cdots,\hat{j})$ or $(\hat{i},\cdots,b_1,a_1,\cdots,\hat{j})$, otherwise the corresponding $\Gamma$ does not contribute.

Each $\Gamma$ can be contained in various BAS amplitudes in the expansion \eref{expan-R2} or \eref{expan-R3}, and each of these BAS amplitudes has the corresponding coefficients $C(\epsilon,\pmb\a_{n-2})$ and $C(\W\epsilon,\pmb\a'_{n-2})$. For a given $\Gamma$ in the order of $\tau^{-t}$, we will prove that the effective part of each corresponding coefficient in \eref{expan-R2} or \eref{expan-R3} behaves as
\bea
C^{\rm eff}(\epsilon,\pmb\a_{n-2})\sim{\cal O}(\tau^c)\,,~~~~C^{\rm eff}(\W\epsilon,\pmb\a'_{n-2})\sim{\cal O}(\tau^{c'})\,,~~~~{\rm with}~c\geq t\,,~c'\geq t\,.~~~~\label{cancel-toprove}
\eea
In other words, the divergence from the propagators can be completely canceled by considering only $C(\epsilon,\pmb\a_{n-2})$ and $\sum_{\pmb\a_{n-2}}$ (or $C(\W\epsilon,\pmb\a'_{n-2})$ and $\sum_{\pmb\a'_{n-2}}$).
The mechanism of such cancellations is as follows.
The first type of cancellation can be traced to kinematic factors like ${\rm tr}({\rm F}_{\pmb\rho})$, ${\rm E}_{\pmb\gamma^f}$ and ${\rm T}^{(i,j)}_{\pmb\pi}$, which serve as building blocks for coefficients in the expansions. If a pair of legs $(a_\ell,b_\ell)$ are also adjacent in one of such $\pmb\rho$, $\pmb\gamma^f$, $\pmb\pi$, the divergence from $1/s_{a_\ell b_\ell}$ is then canceled by the corresponding kinematic factor, due to the definitions of these factors and the re-parameterized kinematic condition \eref{condi-tau}.
If $a_\ell$ and $b_\ell$ are not adjacent in any of these ordered sets, or, they are separated into different ordered sets,
there are two manners of cancellations. The first one, $a_\ell$ and $b_\ell$ appear in one of ${\rm E}_{\pmb\gamma^f}$ as $r_1=a_\ell,Y_{r_1}=k_{b_\ell}$ or $r_1=b_\ell,Y_{r_1}=k_{a_\ell}$, where $r_1$ denotes the first element in $\pmb\gamma^f$, namely $\pmb\gamma^f=\{r_1,\cdots\}$. The divergence from $1/s_{a_\ell b_\ell}$ is then canceled by such ${\rm E}_{\pmb\gamma^f}$. The second manner is,
$\Gamma^{\rm BAS}_n(\cdots,a_\ell,b_\ell,\cdots|\pmb\sigma_n)$ and $\Gamma^{\rm BAS}_n(\cdots,b_\ell,a_\ell,\cdots|\pmb\sigma_n)$ (or, $\Gamma^{\rm BAS}_n(\pmb\sigma_n|\cdots,a_\ell,b_\ell,\cdots)$ and $\Gamma^{\rm BAS}_n(\pmb\sigma_n|\cdots,b_\ell,a_\ell,\cdots)$) cancel each other as in \eref{swapp-pq}, leaving an effective coefficient of a higher order. As in \eref{swapp-pq}, the notation $\Gamma^{\rm BAS}_n(\pmb\sigma'_n|\pmb\sigma_n)$ encodes the contribution from $\Gamma$ to the BAS amplitude ${\cal A}^{\rm BAS}_n(\pmb\sigma'_n|\pmb\sigma_n)$. The above manner of cancellation is then implied by the summation for amplitudes ${\cal A}^{\rm BAS}_n(\cdots,a_\ell,b_\ell,\cdots|\pmb\sigma_n)$ and ${\cal A}^{\rm BAS}_n(\cdots,b_\ell,a_\ell,\cdots|\pmb\sigma_n)$ in the expansion \eref{expan-R2} or \eref{expan-R3}.

By employing the observation \eref{cancel-toprove}, we find
\bea
&&C^{\rm eff}(\epsilon,\pmb\a_{n-2})\,C^{\rm eff}(\W\epsilon,\pmb\a'_{n-2})\,\Gamma^{\rm BAS}_n(\hat{i},\pmb\a_{n-2},\hat{j}|\hat{i},\pmb\a'_{n-2},\hat{j})\,\sim\,{\cal O}(\tau^{c+c'-t})\,,\nn
&&~~~~{\rm with}~c+c'-t>0\,,~{\rm for}~\forall\,t>0\,,~~\label{deduce}
\eea
which holds for any $\Gamma$ contributing $\tau^{-t}$. Therefore, in the expansion \eref{expan-R2} or \eref{expan-R3}, the contribution from any diagram which contains divergent propagators vanishes in the limit $\tau\to 0$. It means one can remove the vanishing kinematic factors without worrying about divergent propagators, and get the reduced effective expansions in \eref{expan-R2-reduced} and \eref{expan-R3-reduced}. The hidden zeros for $R^2$ and $R^3$ amplitudes are then ensured by the KK relation, as discussed earlier.

In subsequent subsections, we will use an explicit example to demonstrate the validity of \eref{cancel-toprove}, and provide a general proof for it. The mechanism of cancellations outlined after \eref{cancel-toprove} will also be elucidated in detail.

\subsection{Cancellations of divergences: $4$-point example}
\label{subsec-cancel-example}

In this subsection, we consider the simplest $4$-point $R^3$ amplitude, to demonstrate the cancellations of divergences.

We choose $\hat{i}=1$, $\hat{j}=3$, $A=2$, $B=4$, then the $\Gamma$ which contains the divergent propagator has only one candidate $\Gamma=1/s_{24}$. The corresponding divergence is at the $\tau^{-1}$ order. Based on the construction for $C_{F^3}(\epsilon,\pmb\a_{n-2})$ described in section \ref{subsec-expan-YMF3}, we can choose the fiducial leg in \eref{YM+2-toYMS} to be $g=\hat{i}=1$, and expand the $4$-point amplitude ${\cal A}^{R^3}_4$ as
\bea
{\cal A}^{R^3}_4=\sum_{\shuffle'}\,C_{F^3}(\W\epsilon,2\shuffle'4)\,&&\Big[{\rm tr}({\rm F}_{24})\,{\cal A}^{\rm YMS}_4(2,4;\underline{1,3}|1,2\shuffle'4,3)\nn
&&+{\rm tr}({\rm F}_{23})\,{\cal A}^{\rm YMS}_4(2,3;\underline{1,4}|1,2\shuffle'4,3)\nn
&&+{\rm tr}({\rm F}_{43})\,{\cal A}^{\rm YMS}_4(4,3;\underline{1,2}|1,2\shuffle'4,3)\nn
&&+{\rm tr}({\rm F}_{243})\,{\cal A}^{\rm YMS}_4(2,4,3;\underline{1}|1,2\shuffle'4,3)\nn
&&+{\rm tr}({\rm F}_{423})\,{\cal A}^{\rm YMS}_4(4,2,3;\underline{1}|1,2\shuffle'4,3)\Big]\,.~~\label{cancel-example-step1}
\eea

The first term on the r.h.s of \eref{cancel-example-step1} corresponds to $\pmb\rho=\{2,4\}$. The re-parameterized kinematic condition \eref{condi-tau}, together with the definition of ${\rm tr}({\rm F}_{\pmb\rho})$ in \eref{defin-F}, imply ${\rm tr}({\rm F}_{24})\sim{\cal O}(\tau^2)$. The factor ${\rm tr}({\rm F}_{24})$ therefore cancels $\tau^{-1}$ from $1/s_{24}$. It is not necessary to further expand ${\cal A}^{\rm YMS}_4(2,4;\underline{1,3}|1,2\shuffle'4,3)$ into ${\cal A}^{\rm BAS}_4(1,2\shuffle4,3|1,2\shuffle'4,3)$, since the cancellation is already manifested. This is an example of the situation that $(a_\ell,b_\ell)$ are adjacent in an ordered set $\pmb\rho$.

The second term with $\pmb\rho=\{2,3\}$ can be further expanded as
\bea
&&{\cal A}^{\rm YMS}_4(2,3;\underline{1,4}|1,2\shuffle'4,3)\nn
&=&(\epsilon_1\cdot k_2)\,{\cal A}^{\rm YMS}_4(2,1,3;\underline{4}|1,2\shuffle'4,3)
+(\epsilon_1\cdot f_4\cdot k_2)\,{\cal A}^{\rm BAS}_4(2,4,1,3|1,2\shuffle'4,3)\nn
&=&(\epsilon_1\cdot k_2)\,(\epsilon_4\cdot k_2)\,{\cal A}^{\rm BAS}_4(2,4,1,3|1,2\shuffle'4,3)+(\epsilon_1\cdot k_2)\,(\epsilon_4\cdot k_{21})\,{\cal A}^{\rm BAS}_4(2,1,4,3|1,2\shuffle'4,3)\nn
&&+(\epsilon_1\cdot f_4\cdot k_2)\,{\cal A}^{\rm BAS}_4(2,4,1,3|1,2\shuffle'4,3)\nn
&=&(\epsilon_1\cdot k_2)\,(\epsilon_4\cdot k_2)\,{\cal A}^{\rm BAS}_4(1,4,2,3|1,2\shuffle'4,3)-(\epsilon_1\cdot k_2)\,(\epsilon_4\cdot k_{21})\,\sum_{\shuffle}\,{\cal A}^{\rm BAS}_4(1,2\shuffle4,3|1,2\shuffle'4,3)\nn
&&+(\epsilon_1\cdot f_4\cdot k_2)\,{\cal A}^{\rm BAS}_4(1,4,2,3|1,2\shuffle'4,3)\,.
\eea
For the first and third terms in the final step, the re-parameterized kinematic condition \eref{condi-tau} forces
\bea
\epsilon_4\cdot k_2\sim{\cal O}(\tau)\,,~~~~~~~~\epsilon_1\cdot f_4\cdot k_2\sim{\cal O}(\tau)\,,
\eea
each of these factors cancels $\tau^{-1}$ from the propagator $1/s_{24}$. These terms are two examples of the manner in which $r_1=b_\ell$ and $Y_{r_1}=k_{a_\ell}$ in the factor ${\rm E}_{\pmb\gamma^f}$. In the second term, the coefficient is independent of the shuffles $\shuffle$
caused by applying the KK relation to ${\cal A}^{\rm BAS}_4(2,1,4,3|1,2\shuffle'4,3)$. Therefore, contributions from $\Gamma=1/s_{24}$ cancel each other as in \eref{swapp-pq} when summing over $\shuffle$. This is an example of the cancellation between two terms, $\Gamma^{\rm BAS}_n(\cdots,a_\ell,b_\ell,\cdots|\pmb\sigma_n)$ and $\Gamma^{\rm BAS}_n(\cdots,b_\ell,a_\ell,\cdots|\pmb\sigma_n)$, which share the same $\Gamma$.
The treatment for the third term with $\pmb\rho=\{4,3\}$ is analogous.

Now we turn to the fourth term on the r.h.s of \eref{cancel-example-step1}, which corresponds to $\pmb\rho=\{2,4,3\}$. The kinematic factor ${\rm tr}(F_{243})$ behaves as ${\rm tr}(F_{243})\sim{\cal O}(\tau)$, as implied by the re-parameterized kinematic condition \eref{condi0-YM}. This factor therefore cancels the divergence from $1/s_{24}$. This is another example where $a_\ell$ and $b_\ell$ are adjacent in an ordered set. The situation of the fifth term with $\pmb\rho=\{4,2,3\}$ is the same.

So far, we have shown that the divergence from the propagator $1/s_{24}$ is completely canceled by considering only $\shuffle$ and $C_{F^3}(\epsilon,2\shuffle 4)$, independently of $\shuffle'$ and $C_{F^3}(\W\epsilon,2\shuffle' 4)$. The effective part $C^{\rm eff}_{F^3}(\epsilon,2\shuffle 4)$
of each $C_{F^3}(\epsilon,2\shuffle 4)$ is of the order $\tau^{c}$ with $c\geq1$. The analogous process shows that the effective part $C^{\rm eff}_{F^3}(\W\epsilon,2\shuffle' 4)$ of each coefficient $C_{F^3}(\W\epsilon,2\shuffle' 4)$ is of the order $\tau^{c'}$, also satisfying $c'\geq1$. Consequently, we have
\bea
C^{\rm eff}_{F^3}(\epsilon,2\shuffle 4)\,C^{\rm eff}_{F^3}(\W\epsilon,2\shuffle' 4)\,\Gamma^{\rm BAS}_4(1,2\shuffle4,3|1,2\shuffle'4,3)\,\sim\,{\cal O}(\tau^{c+c'-1})\,,~~~~{\rm where}~\Gamma={1\over s_{24}}\,,
\eea
which serves as an example of \eref{cancel-toprove}.
The above term vanishes in the limit $\tau\to0$.

\subsection{Cancellations of divergences: general proof}
\label{subsec-cancel-proof}

In this subsection, we provide a general proof for the cancellations of divergences and the effective coefficients in \eref{cancel-toprove}.

As discussed in subsection \ref{subsec-cancel-mecha}, if $\Gamma$ contains $1/s_{ab}$ where $a\in A$ and $b\in B$, then the left and right orderings of each BAS amplitude including $\Gamma$ should take the form $(\cdots,a,b,\cdots)$ or $(\cdots,b,a,\cdots)$. Thus, we can focus on such special pair $(a,b)$ at first, and consider the emergence of the ordering $(\cdots,a,b,\cdots)$ or $(\cdots,b,a,\cdots)$. One need not to worry about other legs adjacent to $a$ or $b$ in the ordering, such like the situation $(\cdots,b',a,b,\cdots)$ and so on, since $1/s_{b'a}$ and $1/s_{ab}$ cannot occur in an individual $\Gamma$ simultaneously.

To see the emergence of $(\cdots,a,b,\cdots)$ or $(\cdots,b,a,\cdots)$, we first consider the left orderings $\pmb\a_{n-2}$ in the expansion \eref{expan-R3} of $R^3$ amplitudes. As described in sections \ref{subsec-expan-YMF3} and \ref{subsec-expan-GRR2R3}, theses orderings $\pmb\a_{n-2}$ are achieved recursively, and the first step is to expand the $R^3$ amplitudes as
\bea
{\cal A}^{R^3}_n=\sum_{\pmb\a'_{n-2}}\,C_{F^3}(\W\epsilon,\pmb\a'_{n-2})\,\Big[&&\sum_{\pmb\rho^{\not{j}},\hat{i}\not\in\pmb\rho^{\not{j}}}\,{\rm tr}({\rm F}_{\pmb\rho^{\not{j}}})\,{\cal A}^{\rm YMS}_n(\pmb\rho^{\not{j}};{\rm G}_n\setminus\rho^{\not{j}}|\hat{i},\pmb\a'_{n-2},\hat{j})\nn
&&+\sum_{\pmb\rho^j,\hat{i}\not\in\pmb\rho^j}\,{\rm tr}({\rm F}_{\pmb\rho^j})\,{\cal A}^{\rm YMS}_n(\pmb\rho^j;{\rm G}_n\setminus\rho^j|\hat{i},\pmb\a'_{n-2},\hat{j})\Big]\,,~~\label{proof-2parts}
\eea
where the fiducial leg is chosen to be $\hat{i}$. In the above, $\pmb\rho^{\not{j}}$ are ordered sets without containing $\hat{j}$,
while $\pmb\rho^j$ are ordered sets with $\hat{j}\in\rho^j$.

For the $\pmb\rho^j$-part, we use the cyclic equivalence to write $\pmb\rho^j=\{\rho_1,\pmb\rho_{|\rho-2|},\hat{j}\}$, and expand each ${\cal A}^{\rm YMS}_n(\pmb\rho^j;{\rm G}_n\setminus\rho^j|\hat{i},\pmb\a'_{n-2},\hat{j})$ as
\bea
&&{\cal A}^{\rm YMS}_n(\pmb\rho^j;{\rm G}_n\setminus\rho^j|\hat{i},\pmb\a'_{n-2},\hat{j})\nn
&=&\sum_{\pmb\gamma^i}\,\sum_{\shuffle}\,{\rm E}_{\pmb\gamma^i}\,{\cal A}^{\rm YMS}_n(\rho_1,\pmb\rho_{|\rho-2|}\shuffle\pmb\gamma^i,\hat{j};{\rm G}_n\setminus\{\rho^j,\gamma^i\}|\hat{i},\pmb\a'_{n-2},\hat{j})\nn
&=&\sum_{\pmb\eta_1}\,\sum_{\pmb\eta_2}\,{\rm E}_{\pmb\gamma^i}\,{\cal A}^{\rm YMS}_n(\pmb\eta_1,\hat{i},\pmb\eta_2,\hat{j};{\rm G}_n\setminus\{\rho^j,\gamma^i\}|\hat{i},\pmb\a'_{n-2},\hat{j})\nn
&=&\sum_{\pmb\eta_1}\,\sum_{\pmb\eta_2}\,(-)^{|\eta_1|}\,{\rm E}_{\pmb\gamma^i}\,\sum_{\shuffle'}\,{\cal A}^{\rm YMS}_n(\hat{i},\pmb\eta_2\shuffle'\pmb\eta^T_1,\hat{j};{\rm G}_n\setminus\{\rho^j,\gamma^i\}|\hat{i},\pmb\a'_{n-2},\hat{j})\nn
&=&\sum_{\pmb\eta_1}\,\sum_{\pmb\eta_2}\,(-)^{|\eta_1|}\,{\rm E}_{\pmb\gamma^i}\,\sum_{\shuffle'}\,\Big[\sum_{\{\pmb\gamma_\ell\}}\,\prod_{\ell}\,\Big(\sum_{\shuffle_\ell}\,{\rm E}_{\pmb\gamma_\ell}\Big)\Big]\,{\cal A}^{\rm BAS}_n(\hat{i},\pmb\eta_2\shuffle'\pmb\eta^T_1\shuffle_1\pmb\gamma_1\shuffle_2\pmb\gamma_2\cdots,\hat{j}|\hat{i},\pmb\a'_{n-2},\hat{j})\,,
~~\label{caserhoj}
\eea
where $\pmb\gamma^i=\{\cdots,\hat{i}\}$, as indicated by the superscript $i$. In the second step, we have relabeled $(\rho_1,\pmb\rho_{|\rho-2|}\shuffle\pmb\gamma^i,\hat{j})$ as $(\pmb\eta_1,\hat{i},\pmb\eta_2,\hat{j})$, and replaced the summations over $\pmb\gamma_i$
and $\shuffle$ by summations over $\pmb\eta_1$ and $\pmb\eta_2$. The third step uses the generalized KK relation \eref{KK-generalized}.
In the final step, each $\{\pmb\gamma_\ell\}$ is an ordered set whose elements are ordered sets $\pmb\gamma_\ell$, satisfying $\gamma_1\cup\gamma_2\cup\cdots={\rm G}_n\setminus\{\rho^j,\gamma^i\}$. The product $\prod_{\ell}\,(\sum_{\shuffle_\ell}\,{\rm E}_{\pmb\gamma_\ell})$ is understood as in \eref{understood-prod}.

Now we can see that, for the $\pmb\rho^j$-part, each left ordering $(\cdots,a,b,\dots)$ has the following origins: (1) $a$ and $b$ are adjacent in one of ordered sets in $\{\pmb\rho^j,\pmb\gamma^i,\pmb\gamma_\ell\}$, and such adjacency is not broken by shuffles; (2) The adjacency of $a$ and $b$ is created by a shuffle in $\{\shuffle,\shuffle',\shuffle_\ell\}$, and is not broken by subsequent shuffles.
As discussed in subsection \ref{subsec-cancel-mecha}, if $a$ and $b$ are adjacent in one of ordered sets in $\{\pmb\rho^j,\pmb\gamma_i,\pmb\gamma_\ell\}$, the corresponding kinematic factor cancels the divergence from $1/s_{ab}$.
On the other hand, if the adjacency of $a$ and $b$ is caused by a shuffle, we should study the cancellation by considering possible shuffles in turn.

Let us start with the first shuffle $\shuffle$ appearing in the first step of \eref{caserhoj}. If the adjacency of $a$ and $b$ is created by this shuffle, there are three situations:
\begin{itemize}
  \item (1) $r_1\not\in\{a,b\}$, where $r_1$ labels the first element in $\pmb\gamma^i$.
      To show the cancellation in this case, the key observation is, the summation over shuffles $\shuffle$ creates $(\cdots,a,b,\cdots)$
      and $(\cdots,b,a,\cdots)$ simultaneously. The kinematic factor ${\rm E}_{\pmb\gamma^i}$ in the current case is invariant under the exchange of $a$ and $b$, thus $\Gamma^{\rm BAS}_n(\cdots,a,b,\cdots)$ and $\Gamma^{\rm BAS}_n(\cdots,b,a,\cdots)$ cancel each other as in \eref{swapp-pq}.
  \item (2) $r_1\in\{a,b\}$, and $q_1\not\in\{a,b\}$, where $q_1$ labels the first element in $\pmb\rho^j$. Again, the shuffles $\shuffle$ generate $(\cdots,a,b,\cdots,)$ and $(\cdots,b,a,\cdots)$ simultaneously. However, in this case, if we swap $a$ and $b$, the kinematic factor ${\rm E}_{\pmb\gamma^i}$ will change, specifically in terms $\cdots f_a\cdot k_b$ or $\cdots f_b\cdot k_a$. For instance, suppose $\pmb\gamma^i=\{a,\hat{i}\}$ and $\pmb\rho^j=\{q_1,b,\hat{j}\}$; then we have ${\rm E}_{a\hat{i}}=\epsilon_{\hat{i}}\cdot f_a\cdot k_{q_1}$ for $(q_1,a,b,\hat{i},\hat{j})$ and ${\rm E}_{a\hat{i}}=\epsilon_{\hat{i}}\cdot f_a\cdot k_{q_1b}$ for $(q_1,b,a,\hat{i},\hat{j})$. These two values of ${\rm E}_{a\hat{i}}$ differ by $\epsilon_{\hat{i}}\cdot f_a\cdot k_b$. Meanwhile, the re-parameterized kinematic condition \eref{condi-tau} implies $(f_a\cdot k_b)^\mu\sim{\cal O}(\tau)$, $(f_b\cdot k_a)^\mu\sim{\cal O}(\tau)$. Therefore, when summing over shuffles, the cancellation between $(\cdots,a,b,\cdots)$ and $(\cdots,b,a,\cdots)$ results in an effective coefficient of a higher order. This higher-order coefficient then cancels the divergence from $1/s_{ab}$.
  \item (3) $\{r_1,q_1\}=\{a,b\}$. In this case, the positions of $a$ and $b$ in the ordering cannot be exchanged, since one of them is fixed at the left end. However, in this case we have $Y_{r_1}=k_a$ or $Y_{r_1}=k_b$, the kinematic factor ${\rm E}_{\pmb\gamma^i}$ then reads $\cdots f_a\cdot k_b$ or $\cdots f_b\cdot k_b$ which behaves as ${\cal O}(\tau)$. Thus, the kinematic factor ${\rm E}_{\pmb\gamma^i}$ cancels the divergence.
\end{itemize}

If the adjacency of $a$ and $b$ arises from a shuffle $\shuffle'$ in the third step of \eref{caserhoj}, then summing over these $\shuffle'$
again yields both $(\cdots,a,b,\cdots)$ and $(\cdots,b,a,\cdots)$ simultaneously. The kinematic factors ${\rm tr}({\rm F}_{\pmb\rho^j})$ and ${\rm E}_{\pmb\gamma^i}$ are clearly independent of these $\shuffle'$, thus $\Gamma^{\rm BAS}_n(\cdots,a,b,\cdots)$ and $\Gamma^{\rm BAS}_n(\cdots,b,a,\cdots)$ again cancel each other.

If the adjacency of $a$ and $b$ arises from one of subsequent shuffles in $\{\shuffle_\ell\}$, the situation is similar as in the first and second cases of $\shuffle$: (1) If $r_1\not\in\{a,b\}$, two terms $\Gamma^{\rm BAS}_n(\cdots,a,b,\cdots)$ and $\Gamma^{\rm BAS}_n(\cdots,b,a,\cdots)$ share the same coefficient therefore cancel each other. (2) If $r_1\in\{a,b\}$, the cancellation between $\Gamma^{\rm BAS}_n(\cdots,a,b,\cdots)$ and $\Gamma^{\rm BAS}_n(\cdots,b,a,\cdots)$ leads to an effective coefficient of a higher order. This effective coefficient cancels the divergence.

So far, we have shown that the divergence from $1/s_{ab}$ is completely canceled in the $\pmb\rho^j$-part in \eref{proof-2parts}. The treatment for the $\pmb\rho^{\not{j}}$-part is similar. We expand the YM$\oplus$BAS amplitudes in the $\pmb\rho^{\not{j}}$-part as
\bea
&&{\cal A}^{\rm YMS}_n(\pmb\rho^{\not{j}};{\rm G}_n\setminus\rho^{\not{j}}|\hat{i},\pmb\a'_{n-2},\hat{j})\nn
&=&\sum_{\pmb\gamma^j_i}\,\sum_{\shuffle}\,{\rm E}_{\pmb\gamma^j_i}\,{\cal A}^{\rm YMS}_n(q_1,\pmb\rho_{|\rho|-2}\shuffle\pmb\gamma^j_i,q_{|\rho|};{\rm G}_n\setminus\{\rho^{\not{j}},\gamma^j_i\}|\hat{i},\pmb\a'_{n-2},\hat{j})\nn
&&+\sum_{\pmb\gamma^j_{\not{i}}}\,\sum_{\shuffle}\,{\rm E}_{\pmb\gamma^j_{\not{i}}}\,{\cal A}^{\rm YMS}_n(q_1,\pmb\rho_{|\rho|-2}\shuffle\pmb\gamma^j_{\not{i}},q_{|\rho|};{\rm G}_n\setminus\{\rho^{\not{j}},\gamma^j_{\not{i}}\}|\hat{i},\pmb\a'_{n-2},\hat{j})\,,~~\label{rho-2parts}
\eea
where $\hat{i}\in\pmb\gamma^j_i$ and $\hat{i}\not\in\pmb\gamma^j_i$. If the adjacency of $a$ and $b$ is caused by a shuffle $\shuffle$ in \eref{rho-2parts}, the cancellation mechanism is the same as that for the shuffle $\shuffle$ in \eref{caserhoj}. That is, we can discuss three cases: (1) $r_1\not\in\{a,b\}$; (2) $r_1\in\{a,b\}$, and $q_1\not\in\{a,b\}$; (3) $\{r_1,q_1\}=\{a,b\}$. The argument and conclusion for each case are essentially the same as those presented previously.

If the adjacency of $a$ and $b$ arises from subsequent shuffles, we need to further expand YM$\oplus$BAS amplitudes in \eref{rho-2parts}. The YM$\oplus$BAS amplitudes in the $\pmb\gamma^j_i$-part share the structure of those in the second line of \eref{caserhoj}; that is, both $\hat{i}$ and $\hat{j}$ are scalars. Meanwhile, the YM$\oplus$BAS amplitudes in the $\pmb\gamma^j_{\not{i}}$-part share the structure of those in the $\pmb\rho^j$-part in \eref{proof-2parts}; i.e., $\hat{j}$ is a scalar while $\hat{i}$ is the gluon. Thus the corresponding cancellations are established by repeating the previous arguments.

We have shown that the divergence from each $1/s_{ab}$ can be eliminated by considering only $\sum_{\pmb\a_{n-2}}$ and $C_{F^3}(\epsilon,\pmb\a_{n-2})$. This conclusion can be directly extended to any $\Gamma\sim{\cal O}(\tau^{-t})$ which contains $1/s_{a_1b_1}\cdots1/s_{a_tb_t}$. Since the divergences never appear in
\bea
\sum_{\pmb\a_{n-2}}\,C_{F^3}(\epsilon,\pmb\a_{n-2})\,\Gamma^{\rm BAS}_n(\hat{i},\pmb\a_{n-2},\hat{j}|\hat{i},\pmb\a'_{n-2},\hat{j})\,,
\eea
we conclude that for any $\Gamma^{\rm BAS}_n(\hat{i},\pmb\a_{n-2},\hat{j}|\hat{i},\pmb\a'_{n-2},\hat{j})\sim{\cal O}(\tau^{-t})$, the effective part of each coefficient behaves as $C^{\rm eff}_{F^3}(\epsilon,\pmb\a_{n-2})\sim{\cal O}(\tau^c)$ with $c\geq t$.
Thus, we have completed the proof of \eref{cancel-toprove}, for $C^{\rm eff}_{F^3}(\epsilon,\pmb\a_{n-2})$.

The proof for $C^{\rm eff}_{\rm YM}(\epsilon,\pmb\a_{n-2})$ is similar and simpler. One can use \eref{YMtoYMS} and \eref{expan-YMS} to expand the $R^2$ amplitudes as
\bea
{\cal A}^{R^2}_n&=&\sum_{\pmb\a'_{n-2}}\,C_{F^3}(\W\epsilon,\pmb\a'_{n-2})\,\Big[\sum_{\pmb\pi}\,{\rm T}^{(\hat{i},\hat{j})}_{\pmb\pi}\,{\cal A}^{\rm YMS}_n(\hat{i},\pmb\pi,\hat{j};{\rm G}_n\setminus\{\hat{i},\hat{j},\pi\}|\hat{i},\pmb\a'_{n-2},\hat{j})\Big]\nn
&=&\sum_{\pmb\a'_{n-2}}\,C_{F^3}(\W\epsilon,\pmb\a'_{n-2})\,\Big[\sum_{\pmb\pi}\,{\rm T}^{(\hat{i},\hat{j})}_{\pmb\pi}\,\Big[\sum_{\{\pmb\gamma_\ell\}}\,\prod_{\ell}\,\Big(\sum_{\shuffle_\ell}\,
{\rm E}_{\pmb\gamma_\ell}\Big)\Big]\nn
&&~~~~~~~~~~~~~~~~~~~~~~~~~~~~{\cal A}^{\rm YMS}_n(\hat{i},\pmb\pi\shuffle_1\pmb\gamma_1\shuffle_2\pmb\gamma_2\cdots,\hat{j}|\hat{i},\pmb\a'_{n-2},\hat{j})\Big]\,.~~\label{proof-CR2}
\eea
Now we consider the divergence from $1/s_{ab}$. If $a$ and $b$ are adjacent in one ordered set in $\{\pmb\pi,\pmb\gamma_\ell\}$, the divergence is canceled by the corresponding kinematic factor. If the adjacency of $a$ and $b$ arises from a shuffle in $\{\shuffle_\ell\}$, then we have: (1) $r_1\not\in\{a,b\}$ where $r_1$ is the first elements in $\pmb\gamma_\ell$. In this case $\Gamma^{\rm BAS}_n(\cdots,a,b,\cdots)$ and $\Gamma^{\rm BAS}_n(\cdots,b,a,\cdots)$ cancel each other. (2) $r_1{\in}\{a,b\}$. In this case the cancellation between $\Gamma^{\rm BAS}_n(\cdots,a,b,\cdots)$ and $\Gamma^{\rm BAS}_n(\cdots,b,a,\cdots)$ gives rise to an effective coefficient at a higher order. This effective coefficient cancels the divergence. The above argument shows that the divergence from $1/s_{ab}$ is always canceled. One can extend this conclusion to arbitrary $\Gamma\sim{\cal O}(\tau^{\-t})$, and conclude \eref{cancel-toprove} for $C^{\rm eff}_{\rm YM}(\epsilon,\pmb\a_{n-2})$.

\section{Summary and discussion}
\label{sec-conclu}

In this paper, we reveal the presence of hidden zeros for higher-derivative YM and GR amplitudes at the tree-level, including the gluon amplitudes with a single insertion of the local $F^3$ operator, as well as $R^2$ and $R^3$ amplitudes on the GR side. By exploiting the universal expansions and the KK relation, these hidden zeros are shown to originate from zeros in BAS amplitudes. The kinematic condition for hidden zeros leads to singular propagators which are unavoidable in unordered graviton amplitudes. We also systematically analyze the cancellations of divergences arising from these singular propagators. These cancellations resolve a key ambiguity in establishing the existence of hidden zeros.

Prior to this work, all amplitudes found to contain hidden zeros consistently exhibited novel factorization behavior called $2$-split near these zeros \cite{Cao:2024qpp}. This leads to an intriguing question: is the presence of hidden zeros always accompanied by---or does it necessarily imply---the $2$-split behavior? The new hidden zeros uncovered in this work offer a perspective for investigating this issue. That is, we can study whether the $F^3$, $R^2$ and $R^3$ amplitudes exhibit $2$-split. This will be the primary objective of our subsequent research.

Another natural and valuable direction for future work is to investigate whether the combination of hidden zeros and factorization on physical poles is sufficient to uniquely determine the $F^3$, $R^2$ and $R^3$ amplitudes. An affirmative answer would open the door to systematically constructing a new effective approach for their calculation, using these zeros as a foundational blueprint---in direct analogy to the novel on-shell recursion relation developed for NLSM amplitudes in \cite{Li:2025suo}.

\section*{Acknowledgments}

We would like to thank Prof. Bo Feng and Liang Zhang for their valuable discussions and cooperation on related research topics. This work is supported by NSFC under Grant No. 11805163.


\appendix

\section{Examples of universal expansions}
\label{sec-example}

In this appendix, we give explicit examples of expansions described in section \ref{sec-expan}.

\subsection{Example of \eref{YM+2-toYMS}}
\label{example-YM+2toYMS}

The first example is the expansion of the $4$-point $F^3$ amplitude ${\cal A}^{F^3}_5(1,2,3,4)$ to YM$\oplus$BAS amplitudes.
We choose the fiducial gluon as $1$. With this choice of the fiducial gluon, the cyclically inequivalent ordered sets $\pmb\rho$ in \eref{YM+2-toYMS} can be
\bea
&&\pmb\rho=\{2,3\}\,,~~~~\pmb\rho=\{2,4\}\,,~~~~\pmb\rho=\{3,4\}\,,\nn
&&\pmb\rho=\{2,3,4\}\,,~~~~\pmb\rho=\{3,2,4\}\,.
\eea
By using the expansion formula \eref{YM+2-toYMS},
the amplitude ${\cal A}^{F^3}_4(1,2,3,4)$ is then expanded as
\bea
{\cal A}^{F^3}_4(1,2,3,4)&=&{\rm tr}({\rm F}_{23})\,{\cal A}^{\rm YMS}_4(2,3;\underline{1,4}|1,2,3,4)+{\rm tr}({\rm F}_{24})\,{\cal A}^{\rm YMS}_4(2,4;\underline{1,3}|1,2,3,4)\nn
&+&{\rm tr}({\rm F}_{34})\,{\cal A}^{\rm YMS}_4(3,4;\underline{1,2}|1,2,3,4)+{\rm tr}({\rm F}_{234})\,{\cal A}^{\rm YMS}_4(2,3,4;\underline{1}|1,2,3,4)\nn
&+&{\rm tr}({\rm F}_{324})\,{\cal A}^{\rm YMS}_4(3,2,4;\underline{1}|1,2,3,4)\,,~~\label{4pt-F3toYMS}
\eea
where the notation $\underline{a,b,c,\cdots}$ labels the unordered set $\{a,b,c,\cdots\}$, such like $\{1,4\}$ in the first term and $\{1,3\}$ in the second term.
The above ${\rm tr}({\rm F}_{\pmb\rho})$ can be evaluated as
\bea
&&{\rm tr}({\rm F}_{23})=\big(f_2\cdot f_3\big)_\mu^{~\mu}\,,~~~~{\rm tr}({\rm F}_{24})=\big(f_2\cdot f_4\big)_\mu^{~\mu}\,,~~~~
{\rm tr}({\rm F}_{34})=\big(f_3\cdot f_4\big)_\mu^{~\mu}\,,\nn
&&{\rm tr}({\rm F}_{234})=-\big(f_2\cdot f_3\cdot f_4\big)_\mu^{~\mu}\,,~~~~{\rm tr}({\rm F}_{324})=-\big(f_3\cdot f_2\cdot f_4\big)_\mu^{~\mu}\,,
~~\label{compu-trF}
\eea
through the rule \eref{defin-F}. According to the definition $f_\ell^{\mu\nu}\equiv k_\ell^\mu\epsilon_\ell^\nu-\epsilon_\ell^\mu k_\ell^\nu$,
one can reduce each ${\rm tr}({\rm F}_{\pmb\rho})$ in \eref{compu-trF} to basic Lorentz invariants of external momenta and polarization vectors,
such as
\bea
&&{\rm tr}({\rm F}_{23})=2\,(k_2\cdot\epsilon_3)\,(k_3\cdot\epsilon_2)-2\,(k_2\cdot k_3)\,(\epsilon_2\cdot\epsilon_3)\,,\nn
&&{\rm tr}({\rm F}_{24})=2\,(k_2\cdot\epsilon_4)\,(k_4\cdot\epsilon_2)-2\,(k_2\cdot k_4)\,(\epsilon_2\cdot\epsilon_4)\,,
\eea
an so on.

\subsection{Example of \eref{expan-YMS}}
\label{subsec-example-5ptYMS}

The next example is the expansion of the $5$-point YM$\oplus$BAS amplitude ${\cal A}^{\rm YMS}_5(1,2,3;\underline{4,5}|1,2,3,4,5)$.
We choose the fiducial leg as $5$, the ordered sets $\pmb\gamma^f$ in \eref{expan-YMS} are then found to be
\bea
\pmb\gamma^5=\{5\}\,,~~~~\pmb\gamma^5=\{4,5\}\,.
\eea
By utilizing the expansion formula \eref{expan-YMS}, the amplitude ${\cal A}^{\rm YMS}_5(1,2,3;\underline{4,5}|1,2,3,4,5)$ is expanded as
\bea
&&{\cal A}^{\rm YMS}_5(1,2,3;\underline{4,5}|1,2,3,4,5)\nn
&=&\sum_{\shuffle}\,(\epsilon_5\cdot Y_5)\,{\cal A}^{\rm YMS}_5(1,2\shuffle 5,3;\underline{4}|1,2,3,4,5)+
\sum_{\shuffle}\,(\epsilon_5\cdot f_4\cdot Y_4)\,{\cal A}^{\rm BAS}_5(1,2\shuffle\{4,5\},3|1,2,3,4,5)\nn
&=&(\epsilon_5\cdot k_1)\,{\cal A}^{\rm YMS}_5(1,5,2,3;\underline{4}|1,2,3,4,5)+(\epsilon_5\cdot k_{12})\,{\cal A}^{\rm YMS}_5(1,2, 5,3;\underline{4}|1,2,3,4,5)\nn
&&+(\epsilon_5\cdot f_4\cdot k_1)\,{\cal A}^{\rm BAS}_5(1,4,5,2,3|1,2,3,4,5)+(\epsilon_5\cdot f_4\cdot k_1)\,{\cal A}^{\rm BAS}_5(1,4,2,5,3|1,2,3,4,5)\nn
&&+(\epsilon_5\cdot f_4\cdot k_{12})\,{\cal A}^{\rm BAS}_5(1,2,4,5,3|1,2,3,4,5)\,,
\eea
where the definition of ${\rm E}_{\pmb\gamma^f}$ in \eref{define-El} is used in the first step, while the definitions of $Y_{r_1}$ and $\sum_{\shuffle}$ are used in the second step.

\subsection{Example of \eref{YMtoYMS}}
\label{subsec-example-5ptYM}

The third example is the expansion of the $4$-point pure YM amplitude ${\cal A}^{\rm YM}_4(1,2,3,4)$ to YM$\oplus$BAS amplitudes. We choose $(i,j)$
in \eref{YMtoYMS} as $i=1$, $j=4$. The proper ordered sets $\pmb\pi$ are found to be
\bea
\pmb\pi=\emptyset\,,~~~~\pmb\pi=\{2\}\,,~~~~\pmb\pi=\{3\},~~~~\pmb\pi=\{2,3\}\,,~~~~\pmb\pi=\{3,2\}\,.
\eea
The expansion formula \eref{YMtoYMS} then leads to
\bea
{\cal A}^{\rm YM}_4(1,2,3,4)&=&(\epsilon_1\cdot\epsilon_4)\,{\cal A}^{\rm YMS}_4(1,4;\underline{2,3}|1,2,3,4)\nn
&&-(\epsilon_1\cdot f_2\cdot\epsilon_4)\,{\cal A}^{\rm YMS}_4(1,2,4;\underline{3}|1,2,3,4)
-(\epsilon_1\cdot f_3\cdot\epsilon_4)\,{\cal A}^{\rm YMS}_4(1,3,4;\underline{2}|1,2,3,4)\nn
&&+(\epsilon_1\cdot f_2\cdot f_3\cdot\epsilon_4)\,{\cal A}^{\rm BAS}_4(1,2,3,4|1,2,3,4)
+(\epsilon_1\cdot f_3\cdot f_2\cdot\epsilon_4)\,{\cal A}^{\rm BAS}_4(1,3,2,4|1,2,3,4)\,,
\eea
which serves as the expansion of the YM amplitude ${\cal A}^{\rm YM}_4(1,2,3,4)$ to YM$\oplus$BAS ones.

\subsection{Expansion of $F^3$ amplitude into BAS KK basis}
\label{example-F3toBAS}

In the final example, we show the expansion of the $4$-point $F^3$ amplitude ${\cal A}^{F^3}_4(1,2,3,4)$ into the BAS KK basis.

We choose $(i,j)=(1,4)$, which are fixed at two ends in the left orderings of BAS amplitudes in the KK basis. By iteratively using the expansion formula \eref{expan-YMS}, one can expand the YM$\oplus$BAS amplitude in the first term on the r.h.s of \eref{4pt-F3toYMS} into such KK basis as
\bea
&&{\cal A}^{\rm YMS}_4(2,3;\underline{1,4}|1,2,3,4)\nn
&=&(\epsilon_4\cdot k_2)\,{\cal A}^{\rm YMS}_4(2,4,3;\underline{1}|1,2,3,4)+(\epsilon_4\cdot f_1\cdot k_2)\,{\cal A}^{\rm BAS}_4(2,1,4,3|1,2,3,4)\nn
&=&(\epsilon_4\cdot k_2)\,{\cal A}^{\rm YMS}_4(3,2,4;\underline{1}|1,2,3,4)+(\epsilon_4\cdot f_1\cdot k_2)\,{\cal A}^{\rm BAS}_4(2,1,4,3|1,2,3,4)\nn
&=&(\epsilon_4\cdot k_2)\,(\epsilon_1\cdot k_3)\,{\cal A}^{\rm BAS}_4(3,1,2,4|1,2,3,4)+
(\epsilon_4\cdot k_2)\,(\epsilon_1\cdot k_{32})\,{\cal A}^{\rm BAS}_4(3,2,1,4|1,2,3,4)\nn
&&+(\epsilon_4\cdot f_1\cdot k_2)\,{\cal A}^{\rm BAS}_4(2,1,4,3|1,2,3,4)\nn
&=&(\epsilon_4\cdot k_2)\,(\epsilon_1\cdot k_3)\,\sum_{\shuffle}\,{\cal A}^{\rm BAS}_4(1,2\shuffle3,4|1,2,3,4)+
(\epsilon_4\cdot k_2)\,(\epsilon_1\cdot k_{32})\,{\cal A}^{\rm BAS}_4(1,2,3,4|1,2,3,4)\nn
&&+(\epsilon_4\cdot f_1\cdot k_2)\,{\cal A}^{\rm BAS}_4(1,2,3,4|1,2,3,4)\,.
\eea
In the second step, we convert the ordering $(2,4,3)$ to $(3,2,4)$ by using the cyclically equivalence. In the final step, we convert all BAS amplitudes to those in the KK basis by utilizing the KK relation \eref{KK} as well as cyclically equivalences.

By performing the similar manipulation, one can expand all YM$\oplus$BAS amplitudes in the remaining terms in \eref{4pt-F3toYMS} into the desired KK basis. Here we present another example, the expansion of the amplitude ${\cal A}^{\rm YMS}_4(2,4;\underline{1,3}|1,2,3,4)$ in the second term on the r.h.s of \eref{4pt-F3toYMS},
\bea
&&{\cal A}^{\rm YMS}_4(2,4;\underline{1,3}|1,2,3,4)\nn
&=&(\epsilon_1\cdot k_2)\,{\cal A}^{\rm YMS}_4(2,1,4;\underline{3}|1,2,3,4)+(\epsilon_1\cdot f_3\cdot k_2)\,{\cal A}^{\rm YMS}_4(2,3,1,4|1,2,3,4)\nn
&=&(\epsilon_1\cdot k_2)\,(\epsilon_3\cdot k_2)\,{\cal A}^{\rm YMS}_4(2,3,1,4|1,2,3,4)+(\epsilon_1\cdot k_2)\,(\epsilon_3\cdot k_{21})\,{\cal A}^{\rm YMS}_4(2,1,3,4|1,2,3,4)\nn
&&+(\epsilon_1\cdot f_3\cdot k_2)\,{\cal A}^{\rm YMS}_4(2,3,1,4|1,2,3,4)\nn
&=&(\epsilon_1\cdot k_2)\,(\epsilon_3\cdot k_2)\,{\cal A}^{\rm YMS}_4(1,3,2,4|1,2,3,4)+(\epsilon_1\cdot k_2)\,(\epsilon_3\cdot k_{21})\,\sum_{\shuffle}\,{\cal A}^{\rm YMS}_4(1,2\shuffle3,4|1,2,3,4)\nn
&&+(\epsilon_1\cdot f_3\cdot k_2)\,{\cal A}^{\rm YMS}_4(1,3,2,4|1,2,3,4)\,.
\eea
By plugging these expansion into \eref{4pt-F3toYMS}, the full expansion of ${\cal A}^{F^3}_4(1,2,3,4)$ into the BAS KK basis is obtained.

\bibliographystyle{JHEP}

\bibliography{reference}

\end{document}